\documentclass[12pt,a4paper]{article}
\usepackage{axodraw}
\usepackage{graphicx}
\usepackage[latin2]{inputenc}
\usepackage{latexsym}
\usepackage{epsf}
\usepackage{epsfig}
\usepackage{cite}

\def\bea{\begin{eqnarray}}  
\def\eea{\end{eqnarray}}  
\def\bc{\begin{center}}
\def\ec{\end{center}}

\begin{document}
\pagestyle{empty}
\begin{flushright}
IFT-01/2006\\
{\tt hep-ph/0601097}\\
{\bf \today}
\end{flushright}
\vspace*{5mm}
\begin{center}

{\large {\bf $Z^\prime$ and the Appelquist-Carrazzone decoupling}}\\
\vspace*{1cm}

{\bf Piotr~H.~Chankowski}, {\bf Stefan~Pokorski}
and {\bf Jakub Wagner}

\vspace{0.5cm}
 
Institute of Theoretical Physics, Warsaw University, Ho\.za 69, 00-681,
Warsaw, Poland

\vspace*{1.7cm}
{\bf Abstract}
\end{center}
\vspace*{5mm}
\noindent
{ 
We consider the electroweak theory with an additional neutral vector boson 
$Z^\prime$ at one loop. We propose a renormalization scheme which makes the 
decoupling of heavy $Z^\prime$ effects manifest. The proposed scheme justifies 
the usual procedure of performing fits to the electroweak data by combining 
the full SM loop corrections to observables with the tree level corrections 
due to the extended gauge structure. Using this scheme we discuss in the model 
with extra an $U(1)^\prime$ group factor 1-loop results for the $\rho$ 
parameters defined in several different ways. 
}
\vspace*{1.0cm}
\date{\today}


\vspace*{0.2cm}
 
\vfill\eject
\newpage

\setcounter{page}{1}
\pagestyle{plain}

\section{Introduction}

For various reasons new physics is expected to show up at the TeV scale.
One of the possibilities, not the least likely one, is that extra gauge 
boson with masses $\sim1$~TeV should be discovered. They are predicted by 
various string inspired models as well as by some models aiming at solving 
the hierarchy problem of the SM. Here 
belong for example Little Higgs models \cite{LITHIG} or models combining 
supersymmetry with the idea of the Higgs doublet as a pseudo-Goldston boson 
\cite{BICHGA,CHFAPOWA}. Before the advent of the LHC, the electroweak data 
are used to constrain parameter spaces of such models. 

The standard methodology used in testing models of new physics against the
electroweak data is that one combines the full one-loop (and also
dominant two-loop) corrections to the relevant observables calculated 
within the SM with modifications stemming from new physics (new gauge 
bosons, new fermions, etc.) accounted at the tree level only. Given that
the top quark mass is known fairly well, this allows to constrain other 
parameters of these models \cite{MASCHST}. 

However, some doubts have been expressed in the literature 
\cite{JEG,CZGLJEZR,DAWSON} about the validity of this standard approach
in models with extended gauge sector. In particular, it has been argued that 
this approach is not valid in theories in which at the tree level $\rho\neq1$ 
since then the entire structure of loop correction is altered and the 
Appelquist-Carrazzone decoupling does not hold. 

To investigate the problem in more detail we consider in this paper the 
simplest extension of the SM with additional $U(1)_E$ gauge group and study 
the one-loop renormalization of the model.\footnote{For earlier 
discussions of the renomalization of the $SU(2)\times U(1)_1\times U(1)_2$ 
models see \cite{DESI,AGMAPEVI}.}
We propose a renormalization 
scheme in which the Appelquist-Carrazzone decoupling is manifest. It combines 
the on-shell renormalization for the three input observables for which we 
conveniently choose $\alpha_{\rm EM}$, $G_F$ and $M_W$ with the 
$\overline{\rm MS}$ scheme for the additional parameters introduced by the 
extended gauge sector. The final expressions for measurable quantities are 
such that
\begin{itemize}
\item they coincide with the SM expression for 
$M_{Z^\prime}\rightarrow\infty$,
\item explicit renormalization scale dependence is only in the 
      $M_{Z^\prime}$ suppressed terms
\item they are are scale independent when the RG running of the parameters 
      is taken into account. Tadpoles play the crucial role here.
\end{itemize}
Our scheme can be contrasted with other renormalization schemes used in the 
literature in which the explicit decoupling of heavy particles ($Z^\prime$)
is lost because also the couplings related to the extended gauge sector 
(couplings of the $U(1)_E$ gauge boson) are expressed in terms of the 
additional to $\alpha_{\rm EM}$, $G_F$ and $M_Z$ (or $M_W$) {\it low energy}
observables like $\sin^2\theta^{\rm eff}_l$ or $\rho$. Our scheme can 
universally be used for $M_{Z^\prime}\sim M_{Z^0}$ or 
$M_{Z^\prime}\gg M_{Z^0}$ whereas the other ones are practical only for 
$M_{Z^\prime}\sim M_{Z^0}$. Indeed, for $M_{Z^\prime}\gg M_{Z^0}$, using e.g.
$\sin^2\theta^{\rm eff}_l$ as an additional input parameter for fixing the
coupling of $Z^\prime$ leads, because of the lack in such a scheme of explicit 
Appelquist-Carrazzone decoupling, to uncertainties which become larger, the 
larger is the $Z^\prime$ mass. The scheme proposed in this paper 
allows to directly constrain by the electroweak data the 
$\overline{\rm MS}$ running parameters of the extended model at a conveniently 
chosen renormalization scale $\mu$, with $\alpha_{\rm EM}$, $G_F$ and $M_W$
chosen as input observables. Furthermore, for $M_{Z^\prime}\gg M_{Z^0}$
it lends justification to the 
standard approach to testing such a model against electroweak data and makes 
it rigorous by specifying {\it what} parameters are being constrained.

As an illustration of the use of our renormalization scheme and in order 
to demonstrate that it leads to explicit Appelquist-Carrazzone decoupling
we clarify various aspects of the $\rho$ parameter(s) in the 
$SU(2)\times U(1)_1\times U(1)_2$ model. First of all, we discuss in detail 
various definitions of $\rho$ and the corresponding tree level results.
Interestingly enough, there exist a definition of $\rho$ in terms of the
low energy neutral to charged current ratio for neutrino processes which
leads to $\rho_{\rm low}=1$ as 
in the SM. Next, we calculate loop corrections to these different 
$\rho$ parameters and show that in the renormalization scheme with explicit 
Appelquist-Carrazzone decoupling the celebrated $m_t^2/m_W^2$ contribution
is always present, The claimed in \cite{JEG,CZGLJEZR} milder, logarithmic 
dependence on $m_t$ is an artifact of a renormalization scheme in which
there is no explicit Appelquist-Carrazzone decoupling.

We also elucidate some specific technical aspects of a theory with
$U(1)_1\times U(1)_2$ group factor related to the mixing of the two 
corresponding gauge bosons resulting in some peculiarities of the RG 
running of the $U(1)$ gauge couplings. 

The plan of the paper is as follows. In Section \ref{sec:generalities} we 
recall the general structure of a $U(1)_1\times U(1)_2$ gauge theory and  
introduce effective charges which allow to cast the Lagrangian in a simple 
form. We express the renormalization group equations for the $U(1)$ couplings 
in terms of these effective couplings. We also introduce the simplest 
extension of the SM by an extra $U(1)$ group factor (with an $SU(2)$ singlet 
scalar vacuum expectation value (VEV) breaking the extra $U(1)$) which will 
serve us as a laboratory to illustrate our main points concerning the loop 
corrections to electroweak observables.
In Section \ref{sec:treerho} we define different $\rho$ parameters, calculate
them at tree level in the model introduced in Section \ref{sec:generalities} 
and show that the leading order contribution of $Z^\prime$ to these parameters 
can be also obtained in the approach using the Appelquist-Carrazzone 
decoupling. In Section \ref{sec:scheme} we define our 
renormalization scheme, and apply it in Section \ref{sec:rho} to calculate 
the corrections to the low energy $\rho$ parameter defined in terms of 
the neutrino processes. In Section \ref{sec:Zmass} we illustrate the interplay
of the proposed scheme with the renormalization group equations derived
in Section \ref{sec:generalities} on the one-loop calculation of the $Z^0$ 
mass. Finally, in Section \ref{sec:Zpeak} we briefly discuss the
calculation of the dominant top bottom contribution to the parameter 
$\rho$ defined in terms of the $Z^0$, $W^\pm$ gauge boson masses and
$\sin^2\theta^\ell_{\rm eff}$ parametrizing the coupling of on-shell $Z^0$
to leptons. Several appendices contain technical details necessary
in the analyzes presented in the main text.

\section{$U(1)_1\times U(1)_2$ gauge theory: couplings and their RG equations}
\label{sec:generalities}

The most general kinetic term for two $U(1)$ gauge fields has the form
\begin{eqnarray}
{\cal L}^{\rm kin}=-{1\over4}f_{\mu\nu}^1f_{\mu\nu}^1
-{1\over4}f_{\mu\nu}^2f_{\mu\nu}^2-{1\over2}\kappa f_{\mu\nu}^1f_{\mu\nu}^2~.
\label{eqn:u1u1_kin}
\end{eqnarray}
$\kappa$ is a real constant constrained by the condition $|\kappa|<1$. The 
most general covariant derivative of a matter field $\psi_k$ is
\begin{eqnarray}
{\cal D}_\mu=\partial_\mu
+i\sum_{a=1}^2\sum_{b=1}^2Y^a_kg_{ab}A^b_\mu~,\label{eqn:covder}
\end{eqnarray}
where the constants $Y^a_k$ play the role of the $U(1)$ charges of $\psi_k$
and $g_{ab}$ are the coupling constants (running couplings in the 
${\overline{\rm MS}}$ renormalization scheme). The gauge transformations 
then are
\begin{eqnarray}
&&A_\mu^a\rightarrow A_\mu^a+\partial_\mu\theta^a~,\nonumber\\
&&\psi_k\rightarrow\exp\left(-i\sum_{a=1}^2\sum_{b=1}^2Y^a_kg_{ab}\theta^b
\right)\psi_k~.\label{eqn:gaugetransf}
\end{eqnarray}
The existence of a whole matrix 
$g_{ab}$ of couplings in place of only one gauge couplings per each $U(1)$
group factor is a peculiarity of the theory with multiple $U(1)$'s 
\cite{AGCOQU,LUXI}. Even if not introduced in the original Lagrangian, the 
last term in (\ref{eqn:u1u1_kin}) and the matrix $g_{ab}$ of couplings are 
generated in the effective action by radiative corrections.

To have simple forms of the tree level propagators, it is convenient to work
in the basis in which the tree-level kinetic mixing is removed.\footnote{It 
is also possible to work with nondiagonal kinetic terms \cite{LUXI,BAKOMARU}.}
By expressing the original $A^{1,2}_\mu$ fields in terms of the new fields 
denoted by $A^Y_\mu$ and $A^E_\mu$ (because they will play the 
roles of the weak hypercharge and extra $U(1)$ gauge bosons, respectively)
\begin{eqnarray}
A^1_\mu={1\over\sqrt{2(1+\kappa)}}A^Y_\mu
       +{1\over\sqrt{2(1-\kappa)}}A^E_\mu~,\phantom{aaaa}
A^2_\mu={1\over\sqrt{2(1+\kappa)}}A^Y_\mu
       -{1\over\sqrt{2(1-\kappa)}}A^E_\mu
\end{eqnarray}
the kinetic cross term disappears (but there will be a 
counterterm $-(1/2)\delta Zf^E_{\mu\nu}f^Y_{\mu\nu}$) and the general form 
(\ref{eqn:covder}) of the covariant derivative does not change. Thus, for 
each matter field $k$ there are charges $Y_k^E$ and $Y_k^Y$ and there are 
four couplings $g_{YY}$, $g_{YE}$, $g_{EY}$, $g_{EE}$. Only three of them 
are independent \cite{AGCOQU}: the $U(1)$ gauge fields can be rotated:
$A^Y=\cos\vartheta\tilde A^Y-\sin\vartheta\tilde A^E$, 
$A^E=\sin\vartheta\tilde A^Y+\cos\vartheta\tilde A^E$, 
without reintroducing the kinetic cross term and such a rotation induces
the corresponding rotations of couplings 
\begin{eqnarray}
\left(\matrix{\tilde g_{YY}\cr\tilde g_{YE}}\right)=
\left(\matrix{\cos\vartheta&\sin\vartheta\cr
-\sin\vartheta&\cos\vartheta}\right)
\left(\matrix{g_{YY}\cr g_{YE}}\right)\phantom{aaaaaa}
\left(\matrix{\tilde g_{EY}\cr\tilde g_{EE}}\right)=
\left(\matrix{\cos\vartheta&\sin\vartheta\cr
-\sin\vartheta&\cos\vartheta}\right)
\left(\matrix{g_{EY}\cr g_{EE}}\right)
\label{eqn:coupltransf}
\end{eqnarray}
The angle $\vartheta$ can be chosen so that one of the four couplings 
vanishes. It is also easy to check that the  following combinations 
\begin{eqnarray}
&&g_{EE}g_{YY}-g_{EY}g_{YE}~,\phantom{aaaaaa}g^2_{EE}+g^2_{EY}~, \nonumber\\
&&g_{YE}g_{EE}+g_{EY}g_{YY}~,\phantom{aaaaaa}g^2_{YY}+g^2_{YE}~,
\label{eqn:invariants}
\end{eqnarray}
are the invariants of the rotations (\ref{eqn:coupltransf}).

The renormalization group equations for the couplings $g_{ab}$ can be 
computed in the standard way \cite{AGCOQU,LUXI} with the result
\begin{eqnarray}
\mu~{d\over d\mu}g_{ba}={1\over16\pi^2}
\sum_{c,d,e} g_{bc}\left[{2\over3}\sum_f(Y_f^dY_f^e)+
{1\over3}\sum_s(Y_s^dY_s^e)\right]~g_{dc}g_{ea}\label{eqn:u1rge}
\end{eqnarray}
where the first sum is over left-chiral fermions and the second one over
complex scalars of the theory. 

As an realistic extension of the SM we consider a theory with the 
$SU(2)_L\times U(1)_Y\times U(1)_E$ electroweak symmetry spontaneously 
broken 
down to $U(1)_{\rm EM}$. The required symmetry breaking is ensured by vacuum
expectation values of an $SU(2)$ doublet $H$ and of a singlet $S$. We assume 
that $S$ is charged under only one $U(1)$, that is $Y_S^Y=0$ 
(but $Y_H^Y\neq0$ 
and $Y^E_H\neq0$), so that $\langle S\rangle = v_S/\sqrt2$ leaves unbroken
$SU(2)_L\times U(1)_Y$. It is then convenient to make the orthogonal field 
redefinition (which does not reintroduce the kinetic mixing term)
\begin{eqnarray}
E_\mu={ g_{EE}A^E_\mu+g_{EY}A^Y_\mu\over\sqrt{g^2_{EE}+g^2_{EY}}}
\phantom{aaaaaaaaaaa}
B_\mu={-g_{EY}A^E_\mu+g_{EE}A^Y_\mu\over\sqrt{g^2_{EE}+g^2_{EY}}}
\label{eqn:redefinition}
\end{eqnarray}
where $E_\mu$ is the combination which becomes massive after $U(1)_E$ 
breaking by $v_S\neq0$ and $B_\mu$ will play the role of the weak 
hypercharge gauge field. The couplings of a generic matter field $\psi_k$ 
to $E_\mu$ and $B_\mu$ are then given by
\begin{eqnarray}
g_y Y_k B_\mu + \left(g_E Y_k^E + g^\prime Y_k^Y\right) E_\mu
\label{eqn:generalcoupling}
\end{eqnarray}
where
\begin{eqnarray}
g_y\equiv{g_{EE}g_{YY}-g_{EY}g_{YE}\over\sqrt{g^2_{EE}+g^2_{EY}}}~,
\phantom{aaaa}
g_E\equiv\sqrt{g^2_{EE}+g^2_{EY}}~,\phantom{aaaa}
g^\prime\equiv{g_{YE}g_{EE}+g_{EY}g_{YY}\over\sqrt{g^2_{EE}+g^2_{EY}}}
\label{eqn:truecouplings}
\end{eqnarray}
are invariants of the transformations (\ref{eqn:coupltransf}). Because only 
three couplings are physical the last invariant, $g^2_{YY}+g^2_{YE}$ in 
(\ref{eqn:invariants}), which does not enter the definitions of $g_y$, 
$g_E$, $g^\prime$ can be expressed in terms of these
\begin{eqnarray}
g^2_{YY}+g^2_{YE}=g_y^2+g^{\prime2}~.\label{eqn:thefourth}
\end{eqnarray}

From (\ref{eqn:generalcoupling}) it follows that $Y_k^Y$ corresponds to the
SM hypercharge. We assume therefore, that the factors  $Y_k^Y$ are as in the 
SM, in particular, $Y^Y_H={1\over2}$. It will also prove convenient to 
introduce effective charges $e_k$ and to rewrite the couplings of matter
fields to the extra gauge boson $E_\mu$ in the form
\begin{eqnarray}
g_Ee_k \equiv g_E Y_k^E+g^\prime Y_k^Y~.\label{eqn:efactdefs}
\end{eqnarray}
With the factors $e_k$ the matter Lagrangian can be written in the naive form 
(frequently used in the literature \cite{APDOHO,CADADOTA}) as if there was no 
mixing of the two $U(1)$ group factors. It is however important to remember 
that $e_k$ are just the way to compactly write the couplings. They
are {\it not} quantum numbers (charges) - except for $e_S$ which is constant.
They {\it do run} with the scale: their RG running can be determined from 
the running of $g_{EE}$, $g_{YY}$, $g_{EY}$, $g_{YE}$ and of $g_E$.

The closed system of the RG equations for the three couplings 
(\ref{eqn:truecouplings}) can 
be readily derived from the general formula (\ref{eqn:u1rge}). Note that these
couplings are defined at any renormalization scale $\mu$ in the (rotating) 
basis in which the kinetic mixing term is absent. Using 
(\ref{eqn:thefourth}) one finds
\begin{eqnarray}
&&{d\over dt}g_E = A^{EE}g_E^3+2A^{EY}g_E^2g^\prime+A^{YY}g_Eg^{\prime2}~,
\nonumber\\
&&{d\over dt}g_y=A^{YY}g_y^3~,\label{eqn:truecouplingsRGE}\\
&&{d\over dt}g^\prime=A^{YY}g^\prime(g^{\prime2}+2g_y^2)
+2A^{EY}g_E(g^{\prime2}+g_y^2)+A^{EE}g_E^2g^\prime~,\nonumber
\end{eqnarray}
where
\begin{eqnarray}
&&A^{ab}={2\over3}\sum_f\left(Y_f^aY_f^b\right)
        +{1\over3}\sum_s\left(Y_s^aY_s^b\right)~.\label{eqn:Afactors}
\end{eqnarray}
With the identification of $Y^Y_k$ as SM hypercharges, the running of $g_y$ 
is exactly as in the SM. This could be expected because of the $U(1)$ Ward 
identity which ensures the absence of threshold corrections to $g_y$
when the heavy massive $E_\mu$ field is decoupled.

In the calculations presented in the following sections we will need RG 
equations for the combinations $e^2_Sg^2_E$ and $e^2_Hg^2_E$ defined by 
(\ref{eqn:efactdefs}). Using (\ref{eqn:truecouplingsRGE}) and 
(\ref{eqn:Afactors}) these RG can be also expressed in terms of the effective 
couplings (\ref{eqn:efactdefs}):
\begin{eqnarray}
{d\over dt}e_S^2g_E^2&=&2e_S^2g_E^2\left({2\over3}\sum_f(e_fg_E)^2
+{1\over3}\sum_s(e_sg_E)^2\right)\nonumber\\
{d\over dt}e^2_Hg^2_E&=&2e^2_Hg^2_E\left({2\over3}\sum_f(e_fg_E)^2
+{1\over3}\sum_s(e_sg_E)^2\right)\label{eqn:RGE}\\
&+&4e_Hg_E\left({2\over3}\sum_fe_fg_EY_f^YY_H^Y
+{1\over3}\sum_se_sg_EY_s^YY_H^Y\right)g_y^2\nonumber
\end{eqnarray}

Finally, we recall the formulae derived in \cite{APDOHO} for gauge boson 
masses appearing as a result of the electroweak breaking by 
$\langle S\rangle=v_S/\sqrt2$ and $\langle H^0\rangle=v_H/\sqrt2$.
The $W^\pm$ boson mass is given as in the SM by $M_W^2={1\over4}g_2^2v_H^2$,
whereas the mass matrix of the neutral gauge bosons in the basis
$(B_\mu,W^3_\mu,E_\mu)$ reads
\begin{eqnarray}
{\cal M}^2_{\rm neut}=
\left(\matrix{
 {1\over4}g^2_yv_H^2 &-{1\over4}g_yg_2v_H^2& {1\over2}g_yg_Ee_Hv_H^2\cr
-{1\over4}g_yg_2v_H^2& {1\over4}g^2_2v_H^2 &-{1\over2}g_2g_Ee_Hv_H^2\cr
 {1\over2}g_yg_Ee_Hv_H^2&-{1\over2}g_2g_Ee_Hv_H^2&g_E^2
 (e_H^2v_H^2+e_S^2v_S^2)}
\right)\label{eqn:neutralgbmassmatrix}
\end{eqnarray}
It is diagonalized by two successive rotations so that the mass eigenstates
are given by 
\begin{eqnarray}
\left(\matrix{B_\mu\cr W^3_\mu\cr E_\mu}\right)=
\left(\matrix{c&-sc^\prime&ss^\prime\cr s&cc^\prime&-cs^\prime\cr
0&s^\prime&c^\prime}\right)
\left(\matrix{A_\mu\cr Z^0_\mu\cr Z^\prime_\mu}\right)~,
\label{eqn:physbosdefs}
\end{eqnarray}
where $c\equiv\cos\theta_W$, $s\equiv\sin\theta_W$ are as in the SM:
$s/c=g_y/g_2$, and $c^\prime\equiv\cos\theta^\prime$, 
$s^\prime\equiv\sin\theta^\prime$, where
\begin{eqnarray}
\tan2\theta^\prime={2(-{1\over2}\sqrt{g_y^2+g_2}g_Ee_Hv_H^2)\over
{1\over4}(g_y^2+g^2_2)v_H^2-g_E^2(e_H^2v_H^2+e_S^2v_S^2)}
\label{eqn:mixingdef}
\end{eqnarray}
The masses of the two gauge bosons, $Z^0$ and $Z^\prime$ are given by
\begin{eqnarray}
M^2_{Z^0}={1\over2}\left(A+B-\sqrt{(A-B)^2+4D^2}\right)~,\nonumber\\
M^2_{Z^\prime}={1\over2}\left(A+B+\sqrt{(A-B)^2+4D^2}\right)~,
\label{eqn:ZZprimemasses}
\end{eqnarray}
where $A=M^2_W/c^2$, $B=e^2_Sg_E^2v_S^2+e^2_Hg_E^2v_H^2$ and 
$D=-(e/2sc)e_Hg_Ev^2_H$. The electric charge $e$ is given by the same formula
as in the SM: $e=g_yc=g_2c$. In \ref{app:massmatrix} we record some 
formulae which will prove indispensable in various manipulations. 

The interactions of the matter fermions with $Z^0$ and $Z^\prime$ bosons 
takes the form
\begin{eqnarray}
{\cal L}_{\rm int}=-J_{Z^0}^\mu Z^0_\mu
-J_{Z^\prime}^\mu Z^\prime_\mu\nonumber
\end{eqnarray}
where the currents are easily found to be
\begin{eqnarray}
J_{Z^0}^\mu&=&\sum_{f=\nu,e,u,d}
\left[{e\over sc}\left(T^3_f-s^2Q_f\right)~
c^\prime+e_fg_E~s^\prime\right]\bar\psi_f\gamma^\mu\mathbf{P}_L\psi_f
\nonumber\\
&+&\sum_{f=e,u,d}
\left[{e\over sc}\left(-s^2Q_f\right)~c^\prime-e_{f^c}g_E~
s^\prime\right]\bar\psi_f\gamma^\mu\mathbf{P}_R\psi_f
\label{eqn:Z0couplings}
\end{eqnarray}
\begin{eqnarray}
J_{Z^\prime}^\mu&=&\sum_{f=\nu,e,u,d}
\left[-{e\over sc}\left(T^3_f-s^2Q_f\right)~
s^\prime+e_lg_E~c^\prime\right]\bar\psi_f\gamma^\mu\mathbf{P}_L\psi_f
\nonumber\\
&+&\sum_{f=e,u,d}
\left[-{e\over sc}\left(-s^2Q_f\right)~s^\prime-e_{f^c}g_E~
c^\prime\right]\bar\psi_f\gamma^\mu\mathbf{P}_R\psi_f~,
\label{eqn:ZPcouplings}
\end{eqnarray}
where $\mathbf{P}_L={1\over2}(1-\gamma^5)$, 
$\mathbf{P}_R={1\over2}(1+\gamma^5)$. The factors in square brackets in
(\ref{eqn:Z0couplings}) and (\ref{eqn:ZPcouplings}) 
define the couplings $c^{Z^0}_{fL,R}$ and $c^{Z^\prime}_{fL,R}$.

Gauge invariance of the Yukawa couplings of the matter fields
\begin{eqnarray}
{\cal L}_{\rm Yuk}=-y_e H^\ast_i l_ie^c -y_t \epsilon_{ij}H_iq_ju^c
-y_d H^\ast_i q_id^c\nonumber
\end{eqnarray}
imposes the conditions (see (\ref{eqn:gaugetransf}))
\begin{eqnarray}
&&Y_{e^c}^a+Y_l^a-Y_H^a=0~,\nonumber\\
&&Y_{u^c}^a+Y_q^a+Y_H^a=0~,\nonumber\\
&&Y_{d^c}^a+Y_q^a-Y_H^a=0~,\nonumber
\end{eqnarray}
where $a=E,Y$. When combined with (\ref{eqn:efactdefs}) they imply
\begin{eqnarray}
&&e_{e^c}+e_l-e_H=0~,\nonumber\\
&&e_{u^c}+e_q+e_H=0~,\label{eqn:erels}\\
&&e_{d^c}+e_q-e_H=0~.\nonumber
\end{eqnarray}

\section{$\rho$ parameters in the $SU(2)_L\times U(1)_Y\times U(1)_E$ model 
and the Appelquist-Carrazzone decoupling}
\label{sec:treerho}

In this section we define various measurable $\rho$ parameters in the 
$SU(2)_L\times U(1)_Y\times U(1)_E$ model and show that at the tree level
the effects of heavy $Z^\prime$ decouple. We then identify the dimension 
six operators which, when added to the SM Lagrangian, reproduce at the tree 
level the leading (in inverse powers of $v_S^2$) corrections to low energy 
observables due to $Z^\prime$.

\subsection{$\rho$ parameters}
\label{susec:rho}

In the SM the measurable parameter $\rho$ can be defined in several different
ways. The simplest is the definition of $\rho$ (call it $\rho_{\rm low}$) as  
the ratio of the coefficients of the neutral and charged current terms in the 
effective low energy four-fermion Lagrangian. Another one is
\begin{eqnarray}
\rho={M^2_W\over M^2_{Z^0}(1-\sin^2\theta)}~,\label{eqn:rho}
\end{eqnarray}
with $\sin^2\theta$ related to measurable quantities in various ways, e.g. 
as the parameter in the on-shell $Z^0$ couplings to fermions as in 
(\ref{eqn:onshellZ0ff}), or by the low energy neutral current Lagrangian for 
e.g. neutrino processes (i.e. as a parameter measuring the admixture of the 
vector-like electromagnetic current in the leptonic weak neutral current in 
the mentioned above low energy four-fermion Lagrangian). 
Finally, $\rho$ (call it $\rho_{Zf}$) can be defined through the coupling of 
on-shell $Z^0$ to fermion-antifermion pairs expressed in terms of the Fermi 
constant measured in the muon decay:
\begin{eqnarray}
{\cal L}_{\rm eff}^{Z^0f\bar f~{\rm on~shell}}
=-\left({\sqrt2G_FM^2_{Z^0}\rho_{Zf}}\right)^{1/2}
\bar\psi_f\gamma^\mu\left(T^3_f-2Q_f\sin^2\theta_{\rm eff}^f 
-T^3_f\gamma^5\right)\psi_fZ^0_\mu~.\label{eqn:onshellZ0ff}
\end{eqnarray}
Independently of the definition used, $\rho=1$ at the tree level due to the 
custodial $SU(2)_V$ symmetry of the SM Higgs potential, Thus, in the SM 
$\rho=1$ is the so-called {\it natural relation}, i.e. the prediction which 
does not depend on the values of the parameters of the model. Of course,
quantum corrections to $\rho$ are numerically different for different 
definitions and do depend on the values of the SM parameters. The usefulness 
of $\rho$ stems from the fact that the dominant contributions (dependent on 
the top quark and Higgs boson masses) to it are universal, that is, the same 
for all definitions of $\rho$. 

Although the different $\rho$ are observables (they are all defined in terms 
of measurable quantities) none of them can be used as an input observable 
in the procedure of renormalization of the SM, just because $\rho=1$ is the 
natural relation.

In the $SU(2)_L\times U(1)_Y\times U(1)_E$ model custodial symmetry is 
broken at the tree level by the $Z^0$-$Z^\prime$ mixing. It is then necessary 
to discuss the analogous $\rho$ parameters in some detail. 
Parameters $\rho$ and $\rho_{Zf}$ can be defined as in the SM, i.e. by the
equations (\ref{eqn:rho}) and (\ref{eqn:onshellZ0ff}), respectively. 
The parameters $\rho_{\rm low}$ is special, because it refers to the specific
form of the low energy effective Lagrangian which needs not be the same as
in the SM. 
In models in which the charged weak currents are unmodified with respect to the
SM the effective Lagrangian for low energy weak interactions takes the 
general form
\begin{eqnarray}
{\cal L}_{\rm eff}=-2\sqrt2G_FJ_+^\mu J_{-\mu}
+{1\over2}\sum_{f_1}\sum_{f_2}\left[
a_{LL}^{f_1f_2}
\left(\bar\psi_{f_1}\gamma^\mu\mathbf{P}_L\psi_{f_1}\right)
\left(\bar\psi_{f_2}\gamma^\mu\mathbf{P}_L\psi_{f_2}\right)\right.\nonumber\\
+a_{RR}^{f_1f_2}
\left(\bar\psi_{f_1}\gamma^\mu\mathbf{P}_R\psi_{f_1}\right)
\left(\bar\psi_{f_2}\gamma^\mu\mathbf{P}_R\psi_{f_2}\right)\nonumber\\
+a_{LR}^{f_1f_2}
\left(\bar\psi_{f_1}\gamma^\mu\mathbf{P}_L\psi_{f_1}\right)
\left(\bar\psi_{f_2}\gamma^\mu\mathbf{P}_R\psi_{f_2}\right)
\label{eqn:lowenefflagr}\\
+a_{RL}^{f_1f_2}\left.
\left(\bar\psi_{f_1}\gamma^\mu\mathbf{P}_R\psi_{f_1}\right)
\left(\bar\psi_{f_2}\gamma^\mu\mathbf{P}_L\psi_{f_2}\right)\right]\nonumber
\end{eqnarray}
where $J_\pm^\mu$ are the standard charged currents. In the SM the second 
part of  (\ref{eqn:lowenefflagr}) can be
rewritten in the form of the product of two neutral currents
\begin{eqnarray}
{\cal L}_{\rm eff}^{\rm NC}=-2\sqrt2G_F J^\mu J_\mu~,
\label{eqn:ncefflagr}
\end{eqnarray}
where 
\begin{eqnarray}
J^\mu=\sum_f\sqrt{\rho_f}~\bar\psi_f\gamma^\mu\left(T^3_f\mathbf{P}_L
-\sin^2\theta_f^{\rm eff}Q_f\right)\psi_f\label{eqn:Jcurrent}
\end{eqnarray}
Moreover, if the fermion mass effects are neglected $\rho_f$ and 
$\sin^2\theta_f^{\rm eff}$ are universal, $\rho_f=\rho$, 
$\sin^2\theta_f^{\rm eff}=\sin^2\theta^{\rm eff}$. $\rho$
can be then factorized out of the neutral current $J^\mu$ and $\rho=1$.

The necessary condition to define the low energy parameter $\rho_f$ (possibly 
dependent on the fermion type) in the 
$SU(2)_L\times U(1)_Y\times U(1)_E$ model is that the second part of
(\ref{eqn:lowenefflagr}) can be written in the current$\times$current form
(\ref{eqn:ncefflagr}). One would then have
\begin{eqnarray}
\sqrt{\rho_{f_1}\rho_{f_2}} =  
-{a_{LL}^{f_1f_2}+a_{RR}^{f_1f_2}-a_{LR}^{f_1f_2}-a_{RL}^{f_1f_2}\over
\sqrt2 G_F2T^3_{f_1}2T^3_{f_2}}
\end{eqnarray}
Computing the diagrams with exchanges of $Z^0$ and $Z^\prime$ between 
the two currents $J^\mu_{Z^0}$ (\ref{eqn:Z0couplings}) and two 
currents $J^\mu_{Z^\prime}$ (\ref{eqn:ZPcouplings}), respectively, and 
exploiting the relations (\ref{eqn:usefulrel1}) and (\ref{eqn:usefulrel2})
it is easy to find 
\begin{eqnarray}
a_{LL}^{f_1f_2}+a_{RR}^{f_1f_2}-a_{LR}^{f_1f_2}-a_{RL}^{f_1f_2}
=-{1\over v^2_H}2T^3_{f_1}2T^3_{f_2}
\phantom{aaaaaaaaaaaaaaaaaaaaaaaa}\nonumber\\
-{(2T^3_{f_1}e_Hg_E+e_{f_1}g_E+e_{f^c_1}g_E)
(2T^3_{f_2}e_Hg_E+e_{f_2}g_E+e_{f^c_2}g_E)\over e_S^2g_E^2v^2_S}
\label{eqn:surprise}
\end{eqnarray}
Due to the relations (\ref{eqn:erels}) the second term vanishes and, since 
at the tree level $1/v_H^2=\sqrt2G_F$, we find (to some surprise) that in 
the $SU(2)_L\times U(1)_Y\times U(1)_E$ model at the tree level 
\begin{eqnarray}
a_{LL}^{f_1f_2}+a_{RR}^{f_1f_2}-a_{LR}^{f_1f_2}-a_{RL}^{f_1f_2}
=-2T^3_{f_1}2T^3_{f_2}\sqrt2G_F\label{eqn:asintheSM}
\end{eqnarray}
as in the SM. However, writing the second part of
(\ref{eqn:lowenefflagr}) in the familiar current$\times$current form is not 
always 
possible. It is only possible, if the following consistency condition holds
\begin{eqnarray}
\left(a_{RR}^{f_1f_2}-a_{LR}^{f_1f_2}\right)
\left(a_{RR}^{f_1f_2}-a_{RL}^{f_1f_2}\right)=-4\sqrt2G_Fa_{RR}^{f_1f_2}
\label{eqn:consistencycond}
\end{eqnarray}
(it follows from the fact that the form (\ref{eqn:ncefflagr}) depends
only on three unknown: $\sqrt{\rho_{f_1}\rho_{f_2}}$, 
$\sin^2\theta_{f_1}^{\rm eff}$ and $\sin^2\theta_{f_2}^{\rm eff}$, whereas
the general form of the second term in (\ref{eqn:lowenefflagr}) has four
independent coefficients). It is straightforward to check that the condition
(\ref{eqn:consistencycond}) is not satisfied in general. 
It is satisfied only by that part of (\ref{eqn:lowenefflagr}) which describes
neutrino reactions. In this case 
$a_{LR}^{f_1\nu_i}=a_{RR}^{f_1\nu_i}=a_{RR}^{\nu_j\nu_i}=0$
and the condition (\ref{eqn:consistencycond}) is trivially satisfied. Thus, 
for neutrino processes one can define the analog of the SM 
$\rho$ parameter as $\rho_{\rm low}\equiv\sqrt{\rho_\nu\rho_f}$
and from (\ref{eqn:surprise}) it follows that at the tree level 
$\rho_{\rm low}=1$ as in the SM. 

In the general case in the $SU(2)_L\times U(1)_Y\times U(1)_E$ 
model even the generalized low energy parameters $\rho_f$ cannot be defined
because the second part of the effective Lagrangian (\ref{eqn:lowenefflagr})
cannot be written in the current$\times$current form. 

It is interesting to contrast $\rho_{\rm low}$ discussed above, for which
$\rho_{\rm low}=1$ at the tree level is a natural relation, with e.g. 
$\rho=M^2_W/M^2_{Z^0}(1-\sin^2\theta)$, with $\sin^2\theta$ identified with
$\sin^2\theta^\ell_{\rm eff}$ in (\ref{eqn:onshellZ0ff}). We find 
\begin{eqnarray}
\sin^2\theta^\ell_{\rm eff}&=&s^2~
{1-{c\over s}e_{\ell^c}{g_E\over e}{s^\prime\over c^\prime}\over
1-2sc~e_H{g_E\over e}{s^\prime\over c^\prime}}
\approx s^2+ s^2\left(2sc~e_H-{c\over s}e_{\ell^c}\right)
{g_E\over e}{s^\prime\over c^\prime}+\dots\nonumber\\
&=&s^2+ \left(s^2~e_H-{1\over 2}e_{\ell^c}\right){e_Hv_H^2\over e_S^2v_S^2}
+\dots\label{eqn:sin2thetaupto}
\end{eqnarray}
where we have used (\ref{eqn:s2c2}).\footnote{Defining 
$\sin^2\theta$ in terms of the structure of the current (\ref{eqn:Jcurrent}) 
for neutrino processes we would get 
\begin{eqnarray}
\sin^2\theta=s^2 +(e_H+e_l)(s^2e_H-{1\over2}e_{e^c}){v_H^2\over e_S^2v_S^2}~.
\nonumber
\end{eqnarray}
}
Using (\ref{eqn:ZZprimemasses}) we get then 
\begin{eqnarray}
\rho\approx
\left(1+{e_H^2g_E^2v_H^2\over e_S^2g_E^2v_S^2}+\dots\right)
\left[1+\left({s^2\over c^2}e_H-{1\over2c^2}e_{\ell^c}\right)
{e_Hv_H^2\over e_S^2v_S^2}+\dots\right]
=1 +{\cal O}\left({v_H^2\over v_S^2}\right).\label{eqn:rhoupto}
\end{eqnarray}
The important difference between $\rho_{\rm low}$ and $\rho$ in the 
$SU(2)_L\times U(1)_Y\times U(1)_E$ model is that the latter does depend on 
some combination of the Lagrangian parameters.\footnote{The fact that at the 
tree level $\rho_{\rm low}=1$ as in the SM makes this observable useless for 
constraining the $SU(2)_L\times U(1)_Y\times U(1)_E$ model as the effects of 
new physics will be always much larger in observables which are modified 
already at the tree level.}
From the above results it is clear that the Appelquist-Carrazzone decoupling
holds at the tree level in the $SU(2)_L\times U(1)_Y\times U(1)_E$ model.
It is also easy to show that it can be easily masked by choosing a low energy
observable like $\sin^2\theta$ (in addition $M_{Z^\prime}$) to fix e.g. 
the coupling $g_E$. To simplify the argument, let us assume that 
$e_{\ell^c}=0$ (at the renormalization scale we are working). Then
$e_H^2v^2_H/e_S^2v_S^2$ in (\ref{eqn:rhoupto}) can be directly expressed
in terms of $\sin^2\theta^\ell_{\rm eff}$ from (\ref{eqn:sin2thetaupto}) 
so that 
\begin{eqnarray}
\rho\approx\left({\sin^2\theta^\ell_{\rm eff}\over s^2}+\dots\right)
\left[1+{\sin^2\theta^\ell_{\rm eff}-s^2\over c^2}+\dots\right]
\end{eqnarray}
and the decoupling is lost!

In the next subsection show the dimension six operators completing the 
SM Lagrangian, which 
reproduce leading terms of the corrections to electroweak observables
found at the tree level.

\subsection{Decoupling at the tree level}

At the tree level the subgroup $U(1)_E$ can be broken independently of the 
breaking of $SU(2)_L\times U(1)_Y$. In this case the gauge field $E_\mu$ 
becomes $Z^\prime$ with a mass $M^2_{Z^\prime}=e^2_Sg_E^2v_S^2$. For $v_S$ 
much higher than the Fermi scale, the electroweak observables can 
be calculated in the $SU(2)_L\times U(1)_Y$ effective theory 
(which is just the SM) supplemented 
with higher dimensional operators generated by decoupling of
heavy $Z^\prime$. This approach yields corrections to the electroweak 
observables due to $Z^\prime$ effects in the form of power series in $1/v_S$.
Below we display the dimension six operators which reproduce the corrections
to different $\rho$ and $\sin^2\theta$ from the preceding subsection
up to ${\cal O}(1/v_S^4)$. 

\begin{figure}[htbp]
\begin{center}
\begin{picture}(320,80)(0,0)
\Photon(10,40)(40,40){3}{4}
\Photon(40,40)(80,40){3}{6}
\Photon(80,40)(110,40){3}{4}
\DashArrowLine(40,40)(20,70){3}
\DashArrowLine(20,10)(40,40){3}
\DashArrowLine(80,40)(100,70){3}
\DashArrowLine(100,10)(80,40){3}
\Vertex(40,40){2}
\Vertex(80,40){2}
\Text(60,30)[]{$Z^\prime$}
\Text(0,47)[]{$W^3$}
\Text(0,33)[]{$B$}
\Text(10,70)[]{$H$}
\Text(10,10)[]{$H$}
\Text(110,70)[]{$H$}
\Text(110,10)[]{$H$}
\Text(120,47)[]{$W^3$}
\Text(120,33)[]{$B$}
\Photon(200,40)(240,40){3}{6}
\Photon(240,40)(270,40){3}{4}
\ArrowLine(200,40)(180,70)
\ArrowLine(180,10)(200,40)
\DashArrowLine(240,40)(260,70){3}
\DashArrowLine(260,10)(240,40){3}
\Vertex(200,40){2}
\Vertex(240,40){2}
\Text(220,30)[]{$Z^\prime$}
\Text(170,70)[]{$f$}
\Text(170,10)[]{$f$}
\Text(270,70)[]{$H$}
\Text(270,10)[]{$H$}
\Text(280,47)[]{$W^3$}
\Text(280,33)[]{$B$}
\end{picture}
\end{center}
\caption{Generating four-fermion operators by the heavy $Z^\prime$.}
\label{fig:otherops}
\end{figure}
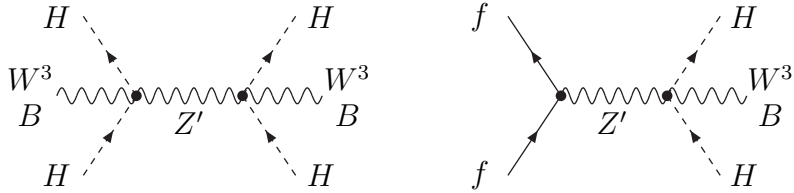

Exchanges of $Z^\prime$ between 
fermion lines are taken into account by adding to the SM Lagrangian
the four-fermion nonrenormalizable operators of the type
\begin{eqnarray}
\Delta{\cal L}_{\rm SM}&=&-{1\over e^2_Sg_E^2v_S^2}
e_l^2g_E^2[\bar\psi_{l_A}\gamma^\mu\mathbf{P}_L\psi_{l_A}]
[\bar\psi_{l_B}\gamma^\mu\mathbf{P}_L\psi_{l_B}]\nonumber\\
&&-{1\over e^2_Sg_E^2v_S^2}
e_l(-e_{e^c})g_E^2[\bar\psi_{e^c_A}\gamma^\mu\mathbf{P}_R\psi_{e^c_A}]
[\bar\psi_{l_B}\gamma^\mu\mathbf{P}_L\psi_{l_B}]\label{eqn:added4Fermi}
\end{eqnarray}
The kinetic term of the electroweak Higgs doublet $H$ gives rise,
through the first diagram of figure \ref{fig:otherops}
to a nonrenormalizable term of the form
\begin{eqnarray}
\Delta{\cal L}_{\rm SM}=-{1\over2}(2e_Hg_E)^2{1\over e^2_Sg_E^2v_S^2}
\left[H^\dagger\left(g_2W^a T^a
+{1\over2}g_yB\right)H\right]^2~.\label{eqn:HHHHeffint}
\end{eqnarray}
Finally, the second diagram shown in figure \ref{fig:otherops}
gives rise the to the interaction:
\begin{eqnarray}
\Delta{\cal L}_{\rm SM}=\sum_f2e_fe_Hg^2_E{1\over e^2_Sg_E^2v_S^2}
\left[H^\dagger\left(g_2W^a_\mu T^a
+{1\over2}g_yB_\mu\right)H\right]\left[\bar f\bar\sigma^\mu f\right]~.
\label{eqn:HHffeffint}
\end{eqnarray}
After the electroweak symmetry breaking the operator (\ref{eqn:HHHHeffint})
gives correction to the $Z^0$ mass squared 
$\Delta M^2_{Z^0}=-(M^2_{Z^0})_{\rm SM}(e_H^2v_H^2/e^2_Sv_S^2)$
whereas the operator (\ref{eqn:HHffeffint}) modify the $Z^0$ couplings
to SM fermions:
\begin{eqnarray}
\Delta{\cal L}_{\rm SM}=-\sum_f{e\over2sc}
{e_fe_H\over e^2_S}{v_H^2\over v_S^2}~
Z^0_\mu\left[\bar f\bar\sigma^\mu f\right]
\approx-\sum_fe_fg_Es^\prime
Z^0_\mu\left[\bar f\bar\sigma^\mu f\right]\nonumber
\end{eqnarray}
they just correspond to terms $e_fg_Es^\prime$ expanded to order $1/v^2_S$ 
in the $Z^0$ couplings (\ref{eqn:Z0couplings}).

At the tree level the three operators (\ref{eqn:added4Fermi}), 
(\ref{eqn:HHHHeffint}) and (\ref{eqn:HHffeffint}) reproduce to order 
$1/M^2_{Z^\prime}\sim1/v_S^2$ all corrections to the low energy 
(compared to $v_S$) observables due to the extended gauge structure
of the model. This is equivalent to the statement that the Appelquist-Carrazone
decoupling works for $Z^\prime$ (at least) at the tree level.

We can illustrate this approach by calculating the corrections due to the 
higher dimensional operators (\ref{eqn:added4Fermi}), (\ref{eqn:HHHHeffint}) 
and (\ref{eqn:HHffeffint}) to the parameter $\rho_{\rm low}$. To this end it 
is sufficient to find the difference $a_{LL}^{\ell\nu}-a_{RL}^{\ell\nu}$ of 
the coefficients in the effective Lagrangian (\ref{eqn:lowenefflagr}). In the 
SM $a_{LL}^{\ell\nu}-a_{RL}^{\ell\nu}=(e^2/4s^2c^2M^2_{Z^0})=1/v_H^2$ and 
since at the tee level $1/v_H^2=\sqrt2G_F$, $\rho_{\rm low}=1$. The corrections
due to the extended gauge structure read
\begin{eqnarray}
(\Delta a_{LL}^{\ell\nu})_{Z^\prime}
=-{1\over g_E^2e^2_Sv^2_S}e_l^2g_E^2~,\phantom{aaaaa}
(\Delta a_{RL}^{\ell\nu})_{Z^\prime}={1\over g_E^2e^2_Sv^2_S}e_le_{e^c}g_E^2
\end{eqnarray}
from the operator (\ref{eqn:added4Fermi}), 
\begin{eqnarray}
(\Delta a_{LL}^{\ell\nu})_{Z^\prime}
={e^2\over4s^2c^2}(1-2s^2){1\over M^2_{Z^0}}
{e_H^2v_H^2\over e_S^2v_S^2}~,\phantom{aaaaa}
(\Delta a_{RL}^{\ell\nu})_{Z^\prime}={e^2\over4s^2c^2}(-2s^2){1\over M^2_{Z^0}}
{e_H^2v_H^2\over e_S^2v_S^2}~,
\end{eqnarray}
from the correction to the $Z^0$ mass produced by the operator 
(\ref{eqn:HHHHeffint}) and
\begin{eqnarray}
(\Delta a_{LL}^{\ell\nu})_{Z^\prime}=-{e^2\over2c^2}{1\over M^2_{Z^0}}
{e_le_Hv_H^2\over e_S^2v_S^2}~,\phantom{aaaaa}
(\Delta a_{RL}^{\ell\nu})_{Z^\prime}
=-{e^2\over4s^2c^2}{1\over M^2_{Z^0}}(2s^2e_l-e_{e^c})
{e_Hv_H^2\over e_S^2v_S^2}~,
\end{eqnarray}
from the correction to the $Z^0$ couplings produced by the operator 
(\ref{eqn:HHffeffint}). Combining these three corrections we find, using the
relations (\ref{eqn:erels}) that 
$\Delta(a_{LL}^{\ell\nu}-a_{RL}^{\ell\nu})=0$.
Other observables can be checked similarly. Subleading 
in $1/v_S$ corrections can be also reproduced upon inclusion in the SM
Lagrangian operators of dimension higher than six.

The equivalence of the two approaches (full calculation versus higher 
dimensional operators) checked above shows that the Appelquist-Carrazzone 
decoupling holds at the tree level. The expectation that it should hold in 
the $SU(2)_L\times U(1)_Y\times U(1)_E$ model to all orders is based 
on the observation that $U(1)_E$ can be broken independently of the breaking 
of $SU(2)_L\times U(1)_Y$. We will propose the scheme which makes it explicit
at one loop and thus show that in particular it is not spoiled by the mixing 
of the gauge fields corresponding to the two $U(1)$ groups.

\section{Renormalization scheme}
\label{sec:scheme}

Before we define our renormalization scheme for the 
$SU(2)_L\times U(1)_Y\times U(1)_E$ extension of the SM, it is instructive
to recall the simplest possible approach to calculating loop corrections
to the electroweak observables within the SM \cite{BAR,POK}. 

Basic (running) parameters of the SM are:\footnote{We denote running 
parameters which are traded for observables by a hat.}
$\hat g_y$, $\hat g_2$ and $\hat v_H$ (or any three other functions of these
parameters, e.g. $\hat\alpha$, $\hat M_Z$ and $\hat s^2$). In the 
renormalization procedure they are expressed in terms of the values of the 
three experimentally measured observables. Traditionally one choses for 
this purpose $G_F$, $\alpha_{\rm EM}$ and $M_Z$. These quantities are  
computed in perturbation calculus using for example the dimensional 
regularization and the $\overline{\rm MS}$ subtraction:
\begin{eqnarray}
&&\alpha_{\rm EM}={\hat g_y^2\hat g_2^2\over4\pi(\hat g_y^2+\hat g_2^2)}
+ \delta\alpha_{\rm EM}={\hat e^2\over4\pi}+\delta\alpha_{\rm EM}
=\hat\alpha+\delta\alpha_{\rm EM}\nonumber\\
&&M_Z^2={1\over4}(\hat g_y^2+\hat g_2^2)\hat v^2=
{1\over4}{\hat e^2\over\hat s^2\hat c^2}\hat v^2+ \delta M^2_Z
 = \hat M_Z^2+ \delta M^2_Z
\label{eqn:basiceqns}\\
&&G_F={1\over\sqrt2\hat v^2}+\delta G_F=
{\hat e^2\over\sqrt24\hat s^2\hat c^2\hat M^2_Z}+\delta G_F
=\hat G_F+\delta G_F \nonumber
\end{eqnarray}
As the corrections $\delta\alpha_{\rm EM}$, $\delta M^2_Z$, $\delta G_F$ 
are calculated in terms of the parameters $\hat\alpha$, $\hat M^2_Z$,
$\hat s^2$ the above relations have to be inverted recursively. At the 
one loop order this is particularly simple: 
\begin{eqnarray}
&&\hat\alpha = \alpha_{\rm EM}-\delta\alpha_{\rm EM}\nonumber\\
&&\hat M_Z^2 =M_Z^2 - \delta M^2_Z\label{eqn:invertbasic}\\
&&\hat G_F= G_F -\delta G_F \nonumber
\end{eqnarray}
where in
$\delta\alpha_{\rm EM}$, $\delta M^2_Z$, $\delta G_F$ one replaces the 
parameters $\hat\alpha$, $\hat M^2_Z$, $\hat s^2$ by $\alpha_{\rm EM}$, $M_Z$ 
and $G_F$ using the tree level relations. 
For any other measurable quantity  ${\cal A}$ we then have
\begin{eqnarray}
&&{\cal A}={\cal A}^{(0)}(\hat\alpha,\hat M^2_Z,\hat G_F)
+ \delta{\cal A}(\hat\alpha,\hat M^2_Z,\hat G_F)+\dots
\end{eqnarray}
where $\delta {\cal A}$ is the one loop contribution to the quantity 
${\cal A}$. This is next written as
\begin{eqnarray}
&&{\cal A}={\cal A}^{(0)}(\alpha_{\rm EM},M^2_Z,G_F)
+ \delta {\cal A}(\alpha_{\rm EM},M^2_Z,G_F)\nonumber\\
&&\phantom{aaaaaaaaaaaaaaaaaaaa}
-{\partial {\cal A}^{(0)}\over\partial \alpha_{\rm EM}}\delta\alpha_{\rm EM}
-{\partial {\cal A}^{(0)}\over\partial M^2_Z}\delta M^2_Z
-{\partial {\cal A}^{(0)}\over\partial G_F}\delta G_F~.\label{eqn:basic}
\end{eqnarray}
The expression (\ref{eqn:basic}) is finite and independent of the 
renormalization scale $\mu$.

The free running parameters of the $SU(2)_L\times U(1)_Y\times U(1)_E$ 
extension of the SM are $g_2$, $v_H$, $v_S$ and the couplings 
$g_{EE}$, $g_{EY}$, $g_{YY}$, $g_{YE}$ (in fact only 
three of them). One way of organizing higher loop calculations in such a model 
is to follow the recipe sketched above and to chose the appropriate number of
input observables, in terms of which one would express all the running 
parameters. 

Clearly, for $M_{Z^\prime}\gg M_{Z^0}$ the parameters of the model form two 
sets: $g_2$, $g_y$ and $v_H$ describe the SM electroweak sector and $v_S$, 
and the remaining gauge couplings describe the $Z^\prime$ sector. 
However, since the 
$Z^\prime$ boson has not yet been discovered and its mass is unknown (assuming 
it exists), the best way to organize loop calculations is such that the 
Appelquist-Carrazzone decoupling (in the case $Z^\prime$ is heavy) would be 
manifest. This condition is not satisfied by schemes in which additional 
parameters related to the heavy particle sector are expressed in terms of 
{\it low energy observables}. Decoupling would be manifest 
if all additional parameters were related to measurable characteristics 
of the heavy particles. Independently of the question of decoupling,
renormalization schemes using the number of observables equal
to the number of free parameters may be difficult to implement in practice
as one has to solve for the running parameters a larger 
set of equations than (\ref{eqn:basiceqns}) in the SM and
the resulting analytical formulae may be very complicated and unmanageable. 

In the fits to the electroweak data,
breakdown of explicit Appelquist-Carrazzone decoupling in a scheme chosen
to compute the  observables may even incorrectly produce {\it upper} bounds
on the additional heavy particles (gauge bosons, Higgs scalars).

In this paper we propose to organize loop calculations into a hybrid scheme
in which the parameters $\hat g_2$, $\hat g_y$ and $\hat v_H$ are expressed 
in terms of $\alpha_{\rm EM}$, $G_F$ and $M_{Z^0}$ (or $M_W$) as in the SM 
and the remaining parameters are kept in the calculations as the
$\overline{\rm MS}$ scheme running parameters. The renormalization scale
$\mu$ for them can be chosen arbitrarily.

As we will show by explicit calculations in the 
$SU(2)_L\times U(1)_Y\times U(1)_E$ model, the advantage of such a hybrid 
scheme\footnote{In fact, such a hybrid scheme is adopted for the usual 
treatment of the strong interaction corrections to the electroweak 
observables: $\hat\alpha_s(\mu)$ is not traded for any observable quantity; 
instead one relies on the fact that the explicit $\mu$ dependence of the 
two-loop contributions should cancel against the $\mu$ dependence of 
$\hat\alpha_s(\mu)$ in one-loop terms.}
is twofold: the Appelquist-Carrazzone decoupling of heavy particle effects is 
made manifest - for heavy particle masses taken to infinity the expressions 
for the observables measured at energies of the order of the electroweak scale 
(or lower) coincide with the SM expression due to the presence of explicit 
suppression by a large mass scale (in the $SU(2)_L\times U(1)_Y\times U(1)_E$ 
model by factors of $1/v_S^2$). Moreover, explicit renormalization scale 
dependence remains only in the terms suppressed by the large mass scale(s). 
The expressions for observables are in fact scale independent when the RG 
running of the parameters is taken into account. Tadpoles play the crucial 
role here \cite{CHWA}. Last but not least, our scheme does not require
solving for running parameters complicated set of equations; in this respect
it is as practical in use as the usual schemes in the SM.

Extensions of the SM are constrained by  precision 
electroweak observables. In our scheme observables are calculated in terms of
$\alpha_{\rm EM}$, $G_F$ and $M_Z$ or $M_W$ (because in the 
$SU(2)_L\times U(1)_Y\times U(1)_E$ model the tree level formula 
(\ref{eqn:ZZprimemasses}) for the $Z^0$ mass is complicated it is much more 
convenient to take as the three input observables $\alpha_{\rm EM}$, $G_F$ 
and $M^2_W$ and compute instead $M^2_{Z^0}$ in terms of these) and the 
additional 
parameters of the model at a conveniently chosen renormalization scale $\mu$.
Fits to the data can then give constraints on these running parameters. 
Moreover, in theories in which the Appelquist-Carrazzone decoupling holds, 
because the loop corrections reduce to their SM form as the heavy mass scale 
is sent to infinity, fairly accurate estimate of the limits imposed by the 
precision data on the additional parameters of the model is possible by 
combining the SM loop corrections with the tree level corrections due to 
``new physics''.

The one loop expressions for the chosen basic  input observables read (see 
\ref{app:input} for details):
\begin{eqnarray}
&&\hat\alpha={\alpha_{\rm EM}\over
1 + \hat{\tilde\Pi}_\gamma(0) -(\hat\alpha/\pi)\ln{\hat M^2_W\over\mu^2}}
\approx\alpha_{\rm EM}\left(
1 - \hat{\tilde\Pi}_\gamma(0) 
+{\alpha_{\rm EM}\over\pi}\ln{M^2_W\over\mu^2}\right)\nonumber\\
&&\hat M_W^2=M^2_W\left(1 -{\hat\Pi_{WW}(M_W^2)\over M_W^2}\right)
\label{eqn:onelooppardet1}\\
&&\hat v_H^2={1\over\sqrt2G_F}(1+\Delta_G)\nonumber
\end{eqnarray}
with $\Delta_G$ given in (\ref{eqn:DeltaG}) and
\begin{eqnarray}
&&\hat s^2={\pi\alpha_{\rm EM}\over\sqrt2G_FM_W^2}(1+\Delta)
\equiv s^2_{(0)}+s^2_{(0)}\Delta\nonumber\\
&&\hat c^2=
{\sqrt2G_FM_W^2-\pi\alpha_{\rm EM}(1+\Delta)\over\sqrt2G_FM_W^2}
\equiv c^2_{(0)}-s^2_{(0)}\Delta\label{eqn:onelooppardet2}
\end{eqnarray}
where
\begin{eqnarray}
\Delta = -\hat{\tilde\Pi}_\gamma(0)
+{\hat\alpha\over\pi}\ln{\hat M^2_W\over\mu^2}
+{\hat\Pi_{WW}(M_W^2)\over M^2_W}+\Delta_G\label{eqn:Delta}
\end{eqnarray}
(as usually $\hat{\tilde\Pi}_\gamma(q^2)$ is defined by 
$\hat\Pi_{\gamma\gamma}(q^2)=q^2\hat{\tilde\Pi}_\gamma(q^2)$, i.e. it is the 
residue of the photon propagator).

Using this scheme we will explicitly demonstrate that in the 
$SU(2)_L\times U(1)_Y\times U(1)_E$ extension of the SM the 
Appelquist-Carrazzone decoupling does hold. To this end we will
compute in our scheme the two different $\rho$ parameters defined 
as in Section 3 in terms
of the observables: $\rho_{\rm low}$ defined by the effective Lagrangian
for $\nu_\mu e^-$ elastic scattering and 
$\rho\equiv M^2_W/M_Z^2(1-\sin^2\theta^\ell_{\rm eff})$ where 
$\sin^2\theta^\ell_{\rm eff}$ parametrizes the effective coupling of on-shell
$Z^0$ to $l^+l^-$ pair. In particular we will demonstrate that the 
celebrated $m^2_t/M^2_W$ term is present in both cases.

\section{Decoupling of $Z^\prime$ effects in  $\rho_{\rm low}$ at 1-loop}
\label{sec:rho}

As an exercise, in order to demonstrate the working of our renormalization
scheme we will compute one loop corrections to the low energy parameter 
$\rho_{\rm low}$ defined by the $\nu_\mu e^-\rightarrow \nu_\mu e^-$ elastic 
scattering. Since $\rho_{\rm low}=1$ at the tree level is a natural relation
in the $SU(2)\times U(1)_Y\times U(1)_E$ model,  the one loop corrections to
$\rho_{\rm low}$ should be finite when $1/v_H^2$ in (\ref{eqn:surprise}) is 
expressed in terms of $G_F$ with one loop accuracy. 

At one loop the direct generation number dependent fermion contribution comes
through the ``oblique'' corrections to $a_{LL}^{e\nu}-a_{RL}^{e\nu}$:
\begin{eqnarray}
(a_{LL}^{e\nu}-a_{RL}^{e\nu})_{\rm 1-loop}&=&
c^{Z^0}_{\nu L}a^{Z^0}_e{1\over M_{Z^0}^2}{\Pi_{Z^0Z^0}(0)\over M_{Z^0}^2}
+c^{Z^\prime}_{\nu L}a^{Z^\prime}_e
{1\over M_{Z^\prime}^2}{\Pi_{Z^\prime Z^\prime}(0)\over M_{Z^\prime}^2}
\nonumber\\
&+&c^{Z^\prime}_{\nu L}a^{Z^0}_e
{1\over M_{Z^0}^2}{\Pi_{Z^0Z^\prime}(0)\over M_{Z^\prime}^2}
+c^{Z^0}_{\nu L}a^{Z^\prime}_e
{1\over M_{Z^0}^2}{\Pi_{Z^0Z^\prime}(0)\over M_{Z^\prime}^2}~,
\label{eqn:Z0ZPtorho}
\end{eqnarray}
where $Z_i$ denotes $Z^0$ or $Z^\prime$, $a^{Z_i}_f=c^{Z_i}_{fL}-c^{Z_i}_{fR}$,
and the couplings $c_{fL,R}^{Z^0}$, 
($c_{fL,R}^{Z^\prime}$) of $Z^0$ ($Z^\prime$) to left and right-chiral leptons 
are defined by (\ref{eqn:Z0couplings}), (\ref{eqn:ZPcouplings}). 
The self energies $\Pi_{Z_iZ_j}$ contain in principle also tadpole
contributions. Another generation number dependent contribution to $\rho$
arises from $\hat\Pi_{WW}(0)/\hat M^2_W$ after expressing $1/\hat v_H^2$ in 
the tree level term (\ref{eqn:surprise}) with one loop accuracy 
\begin{eqnarray}
(a_{LL}^{e\nu}-a_{RL}^{e\nu})_{\rm tree}=
{1\over\hat v_H^2}=\sqrt2G_F(1-\Delta_G)\label{eqn:aminusbtree}
\end{eqnarray}
with $\Delta_G$ given by (\ref{eqn:DeltaG}).
\vskip0.3cm

\noindent {\it Fermionic contribution to $\rho_{\rm low}$}

\vskip0.2cm
\noindent The top-bottom quark contribution to 1-particle irreducible part of
$\hat\Pi_{WW}$ is the same as in the SM
\begin{eqnarray}
\hat\Pi_{WW}(0)={\hat e^2\over\hat s^2}N_c\left[2\tilde A(0,m_t,m_b)
-{1\over2}(m_t^2+m_b^2)b_0(0,m_t,m_b)\right]~,\label{eqn:PiWW0}
\end{eqnarray}
where $N_c=3$. The 1-particle irreducible part of $\hat\Pi_{Z_iZ_j}(0)$ 
can be simplified to 
\begin{eqnarray}
\hat\Pi_{Z_iZ_j}(0)=-2a^{Z_i}_ta^{Z_j}_tN_cm_t^2b_0(0,m_t,m_t)
-2a^{Z_i}_ba^{Z_j}_bN_cm_b^2b_0(0,m_b,m_b)\nonumber
\end{eqnarray}
Contributions of the other fermion fermions can be written analogously.
When inserted into (\ref{eqn:Z0ZPtorho}) the fermion $f$ contribution to
$\hat\Pi_{Z_iZ_j}(0)$ factorizes
\begin{eqnarray}
(a_{LL}^{e\nu}-a_{RL}^{e\nu})_{\rm 1-loop}^{(f)}=
-\left({a^{Z^0}_ea^{Z^0}_f\over M^2_{Z^0}}+
{a^{Z^\prime}_ea^{Z^\prime}_f\over M^2_{Z^\prime}}\right)
\left({c^{Z^0}_{\nu L}a^{Z^0}_f\over M^2_{Z^0}}+
{c^{Z^\prime}_{\nu L}a^{Z^\prime}_f\over M^2_{Z^\prime}}\right)~
2m_f^2N_c~b_0(0,m_f,m_f)
\nonumber
\end{eqnarray}
and computing the factors in brackets using  (\ref{eqn:Z0couplings}), 
(\ref{eqn:ZPcouplings}) and the formulae (\ref{eqn:usefulrel1}),
(\ref{eqn:usefulrel2}) one finds (omitting $1/16\pi^2$)
\begin{eqnarray}
(a_{LL}^{e\nu}-a_{RL}^{e\nu})_{\rm 1-loop}^{t,b}=
{2\over v_H^4}m^2_tN_c\ln{m^2_t\over\mu^2}\times
\left[1-{v^2_H\over v_S^2}
{(e_l+e_{e^c}-e_H)(e_q+e_{u^c}+e_H)\over e_S^2}\right]\nonumber\\
\times\left[1+{v^2_H\over v_S^2}
{(e_l+e_H)(e_q+e_{u^c}+e_H)\over e_S^2}\right]\phantom{aaaai}\nonumber\\
+{2\over v_H^4}m^2_bN_c\ln{m^2_b\over\mu^2}\times
\left[1+{v^2_H\over v_S^2}
{(e_l+e_{e^c}-e_H)(e_q+e_{d^c}-e_H)\over e_S^2}\right]\nonumber\\
\times\left[1-{v^2_H\over v_S^2}
{(e_l+e_H)(e_q+e_{d^c}-e_H)\over e_S^2}\right]\phantom{aaaai}\nonumber
\end{eqnarray}
The first terms in square brackets reproduce the SM contribution. 
The other terms are simply zero due to the relations (\ref{eqn:erels}).
Combining this with the top bottom contribution to 
$\hat\Pi_{WW}(0)$ in (\ref{eqn:aminusbtree}) one finds that the fermionic
``oblique'' contribution to $\rho_{\rm low}$ is finite and exactly reproduces 
the one-loop SM result
\begin{eqnarray}
\Delta\rho_{\rm low}={N_c\over16\pi^2}\sqrt2G_Fg(m_t,m_b)+\dots=
{N_c\over16\pi^2}\sqrt2G_Fm^2_t+\dots
\end{eqnarray}
(the function $g(m_1,m_2)$ is defined in \ref{app:functions}). Thus, we 
explicitly demonstrate that in the $SU(2)_L\times U(1)_Y\times U(1)_E$ model
the celebrated $\propto m^2_t$ contribution is present in the $\rho$ 
parameter defined in terms of low energy neutrino processes.
\vskip0.3cm

\noindent {\it Bosonic contribution $\rho_{\rm low}$}

\vskip0.2cm
\noindent 
The circumstance simplifying calculation of the the vertex and self energy 
corrections to external lines to the $\nu_\mu e^-\rightarrow\nu_\mu e^-$ 
amplitude is that (due to the 
corresponding $U(1)$ Ward identities) the corrections to the vertices due 
to virtual $Z^0$ and $Z^\prime$ are exactly canceled by the virtual $Z^0$ 
and $Z^\prime$ contributions to the self energies. For the corrections due 
to virtual $W$ one finds 
\begin{eqnarray}
16\pi^2\left(a_{LL}^{e\nu}\right)_{\rm1-loop}^{\rm vert}=\left[
c^{Z^0}_{eL}{1\over M^2_{Z^0}}
\left(\hat e^3{\hat c\over\hat s^3}c^\prime\right)
+c^{Z^\prime}_{eL}{1\over M^2_{Z^\prime}}
\left(-\hat e^3{\hat c\over\hat s^3}s^\prime\right)
\phantom{aaaaaaaaaa}
\right.\nonumber\\
\left.
+c^{Z^0}_{\nu L}{1\over M^2_{Z^0}}
\left(-\hat e^3{\hat c\over\hat s^3}c^\prime\right)
+c^{Z^\prime}_{\nu L}{1\over M^2_{Z^\prime}}
\left(\hat e^3{\hat c\over\hat s^3}s^\prime\right)
\right]\left(\eta_{\rm div}+\ln{\hat M^2_W\over\mu^2}\right)
\end{eqnarray}
\begin{eqnarray}
16\pi^2\left(a_{RL}^{e\nu}\right)_{\rm1-loop}^{\rm vert}=\left[
c^{Z^0}_{eR}{1\over M^2_{Z^0}}
\left(\hat e^3{\hat c\over\hat s^3}c^\prime\right)
+c^{Z^\prime}_{eR}{1\over M^2_{Z^\prime}}\left(
-i\hat e^3{\hat c\over\hat s^3}s^\prime\right)\right]
\left(\eta_{\rm div}+\ln{\hat M^2_W\over\mu^2}\right)
\end{eqnarray}
and, after using the relations (\ref{eqn:usefulrel1}), (\ref{eqn:usefulrel2}), 
\begin{eqnarray}
16\pi^2\left(a_{LL}^{e\nu}-a_{RL}^{e\nu}\right)_{\rm1-loop}^{\rm vert}=
-{4\over\hat v_H^2}\hat e^2{\hat c^2\over\hat s^2}\left[
1+{1\over2}{\hat v^2_H\over\hat v^2_S}{2e_H^2-e_He_{e^c}\over e^2_S}\right]
\left(\eta_{\rm div}+\ln{\hat M^2_W\over\mu^2}\right)~.
\label{eqn:vertexcontr}
\end{eqnarray}
Using relations (\ref{eqn:usefulrel1}), (\ref{eqn:usefulrel2}) and the 
results for $\hat\Pi_{\gamma Z^0}(0)$ and $\hat\Pi_{\gamma Z^\prime}(0)$ 
which can be extracted from \ref{app:alphaEM} one can also check that the 
potentially singular at zero momentum transfer ``oblique'' corrections to 
the $\nu_\mu e\rightarrow\nu_\mu e$ scattering amplitude cancel against 
the singular contribution of the photon exchange between the tree level
$ee\gamma$ and one loop $\nu\nu\gamma$ vertices as in the SM \cite{POK}.

The bosonic contribution to (\ref{eqn:Z0ZPtorho}) can be calculated using 
the formulae collected in \ref{app:VVself}. The structure of the $W^+W^-$, 
$G^\pm W^\mp$, $G^+G^-$, $G^0h^0$ and $G^\prime S^0$ contribution to 
$\Pi_{Z_iZ_j}$ is such that they can be written in the form
\begin{eqnarray}
\Pi_{Z_iZ_j}^{(k)}(q^2)=\alpha_{Z_i}^{(k)}\alpha_{Z_j}^{(k)}\Pi^{(k)}(q^2)
\label{eqn:factform}
\end{eqnarray}
which when used in the $e\nu\rightarrow e\nu$ amplitude leads to the 
factorization observed already for the fermionic contribution:
\begin{eqnarray}
a^{e\nu}_{LL}=
\left({c_{\nu L}^{Z^0}\alpha_{Z^0}^{(k)}\over M^2_{Z^0}}+
{c_{\nu L}^{Z^\prime}\alpha_{Z^\prime}^{(k)}\over M^2_{Z^\prime}}\right)
\left({c_{eL}^{Z^0}\alpha_{Z^0}^{(k)}\over M^2_{Z^0}}+
{c_{eL}^{Z^\prime}\alpha_{Z^\prime}^{(k)}\over M^2_{Z^\prime}}\right)
\Pi^{(k)}(q^2)\label{eqn:factorization}
\end{eqnarray}
and similarly for $a^{e\nu}_{RL}$. This allows to easily calculate the 
divergent part of the corresponding contributions to 
$a_{LL}^{e\nu}-a_{RL}^{e\nu}$ (of these only $W^+W^-$ and $G^\pm W^\mp$
are divergent). Using the tricks 
(\ref{eqn:usefulrel1}), (\ref{eqn:usefulrel2}) and (\ref{eqn:erels}) it is
\begin{eqnarray}
{1\over\hat v_H^2}2\hat e^2{\hat c^4\over\hat s^2}
\left[1 + {v^2_H\over v^2_S}{e_H(e_H+e_l)\over e^2_S}\right]
-{\hat e^2\over\hat v_H^2}\left(2\hat s^2 -2 \hat c^2{v^2_H\over v^2_S}
{e_H(e_H+e_l)\over e^2_S}\right)\eta_{\rm div}~.
\label{eqn:WWobldiv}
\end{eqnarray}
The divergences of the $Z^0h^0$ and $Z^\prime h^0$  loop  
contributions to $\Pi_{Z_iZ_j}$ can be combined to yield
\begin{eqnarray}
\left[\Pi_{Z_iZ_j}\right]_{\rm div}=\alpha_{Z_i}\alpha_{Z_j}
\left({\hat e^2\over\hat s^2\hat c^2}+4e_H^2g_E^2\right){\hat v_H^2\over4}
\eta_{\rm div}\nonumber
\end{eqnarray}
with $\alpha_{Z^0}=-{\hat e\over\hat s\hat c}c^\prime +2e_Hg_Es^\prime$
and $\alpha_{Z^\prime}={\hat e\over\hat s\hat c}c^\prime +2e_Hg_Es^\prime$.
The corresponding divergent contributions to $a_{LL}^{e\nu}-a_{RL}^{e\nu}$ 
is then
\begin{eqnarray}
-{1\over\hat v_H^2}
\left({\hat e^2\over\hat s^2\hat c^2}+4e_H^2g_E^2\right)\eta_{\rm div}
\label{eqn:Zhobldiv}
\end{eqnarray}
The other ``oblique'' bosonic contributions are finite. 
It is also easy to check that the tadpole contributions to the vector
boson self energies cancel out in the difference 
$a^{e\nu}_{LL}-a^{e\nu}_{RL}$.

Finally we record for completeness the finite contributions of the box 
diagrams to the coefficients $a_{LL}^{\nu e}$ and $a_{LR}^{\nu e}$ of the 
low energy Lagrangian (\ref{eqn:lowenefflagr}). We find
\begin{eqnarray}
16\pi^2a_{LL}^{\nu e}&=&
 {1\over M^2_{Z^0}}3(c_{\nu L}^{Z^0})^2(c_{eL}^{Z^0})^2
+{1\over M^2_{Z^\prime}}3(c_{\nu L}^{Z^\prime})^2(c_{eL}^{Z^\prime})^2
\nonumber\\
&+&{1\over M^2_{Z^\prime}-M^2_{Z^0}}\ln\left({M^2_{Z^\prime}\over M^2_{Z^0}}
\right)
6~c_{\nu L}^{Z^0}c_{\nu L}^{Z^\prime} c_{eL}^{Z^0}c_{eL}^{Z^\prime}
+{\hat e^4\over\hat s^4M^2_W}\\
16\pi^2a_{LR}^{\nu e}&=&
-{1\over M^2_{Z^0}}3(c_{\nu L}^{Z^0})^2(c_{eR}^{Z^0})^2
-{1\over M^2_{Z^\prime}}3(c_{\nu L}^{Z^\prime})^2(c_{eR}^{Z^\prime})^2
\nonumber\\
&-&{1\over M^2_{Z^\prime}-M^2_{Z^0}}\ln\left({M^2_{Z^\prime}\over M^2_{Z^0}}
\right)
6 c_{\nu L}^{Z^0}c_{\nu L}^{Z^\prime} c_{eR}^{Z^0}c_{eR}^{Z^\prime}\nonumber
\end{eqnarray}
From these formulae the box contribution to $\rho_{\rm low}$ can be easily
obtained.

Combining the results (\ref{eqn:vertexcontr}), (\ref{eqn:WWobldiv}),
(\ref{eqn:Zhobldiv}) with the divergent part of $\Delta_G$ in 
(\ref{eqn:aminusbtree}) given by (\ref{eqn:boxesandvertices}) and 
(\ref{eqn:WWat0div}) one easily finds that the total one loop contribution
to the $\rho_{\rm low}$ parameter defined in terms of the 
$\nu e\rightarrow\nu e$ scattering amplitude is finite and, since the
coefficient of $\ln(1/\mu^2)$ is the same as that of $\eta_{\rm div}$,
independent of the renormalization scale. Moreover, it is easy to see, that 
in the limit $v_S\rightarrow\infty$ one recovers the SM result i.e. the 
Appelquist-Carrazzone decoupling is manifest. 

If $\sin^2\theta^{\rm eff}_\ell$
is used as an additional observable, the explicit decoupling is lost. 
This is because one has then to express $g_E$ and $v_S$ in the one-loop 
contribution through $M_{Z^\prime}$ and  $\sin^2\theta^{\rm eff}_\ell$
(to zeroth order accuracy) with the effect already described:
the explicit suppression factor $\propto1/v^2_S$ is then replaced by the
difference of $\sin^2\theta^{\rm eff}_\ell-s^2_{(0)}$ which is finite
and does not vanish as $v_S\rightarrow\infty$.

\section{One loop calculation of $M_{Z^0}^2$}
\label{sec:Zmass}

In this section we compute in our scheme $M_{Z^0}^2$. Unlike the previous
example of $\rho_{\rm low}$, the tree level formula for $M_{Z^0}^2$ does
depend on the parameters of the extended gauge sector. Therefore, in the 
one loop result for $M_{Z^0}^2$ in our scheme explicit dependence on the 
renormalization scale $\mu$ will remain. We will however
show that the conditions for the heavy $Z^\prime$ effects to decouple 
are satisfied: the part of the result which does not vanish as 
$v_S\rightarrow\infty$ is independent of $\mu$ and takes the SM form.
Furthermore, we will show that the whole result for $M_{Z^0}^2$ is 
independent of the renormalization scale if the dependence on $\mu$
of the parameters in the zeroth-order expression is taken into account.
This  constitutes a nontrivial check of the renormalization group 
equations (\ref{eqn:truecouplingsRGE})-(\ref{eqn:RGE}) and of our
renormalization scheme.

We calculate now the one loop corrections 
to the $Z^0$ boson mass. It is given by the formula\footnote{Mixing of
$Z^0$ with $Z^\prime$ is formally a two-loop effect.}
\begin{eqnarray}
M_{Z^0}^2=\hat M_{Z^0}^2 + \Pi_{Z^0Z^0}(M_{Z^0}^2)\nonumber
\end{eqnarray}
where the tree level term $\hat M_{Z^0}^2$ is given by 
(\ref{eqn:ZZprimemasses}). 
The running parameters $\hat e$, $\hat s$, $\hat c$, $\hat v_H$ in 
$\hat M_{Z^0}^2$ have to be expressed in terms of the input
observables $G_F$, $M_W^2$ and $\alpha_{\rm EM}$ with one loop accuracy by 
using the relations (\ref{eqn:onelooppardet1}), (\ref{eqn:onelooppardet2}).
This gives 
\begin{eqnarray}
&&A_0+\delta A={M^2_W\over c^2_{(0)}}\left\{1-{\hat\Pi_{WW}(M_W^2)\over M^2_W}
+{s^2_{(0)}\over c^2_{(0)}}\Delta\right\}\nonumber\\
&&B_0+\delta B=g^2_Ee_S^2v_S^2+{g^2_Ee_H^2\over\sqrt2G_F}
\left(1+\Delta_G\right)\label{eqn:ABDdeltaABD}\\
&&D_0+\delta D=-{1\over2}e_Hg_E{e_{(0)}\over s_{(0)}c_{(0)}\sqrt2G_F}
\left\{1+{1\over2}{s^2_{(0)}\over c^2_{(0)}}\Delta 
-{1\over2}{\hat\Pi_{WW}(M_W^2)\over M^2_W}
+{1\over2}\Delta_G\right\}\nonumber
\end{eqnarray}
where $e_{(0)}\equiv\sqrt{4\pi\alpha_{\rm EM}}$ and $\Delta$ and $\Delta_G$
are given by (\ref{eqn:Delta}) and (\ref{eqn:DeltaG}), respectively.
In agreement with the prescription (\ref{eqn:basic}) we then have 
$2M^2_{Z^0}=2(M^2_{Z^0})_{(0)}+2\delta M^2_{Z^0}$ where $(M^2_{Z^0})_{(0)}$
is given by (\ref{eqn:ZZprimemasses}) with $A$, $B$, $D$ replaced by 
$A_0$, $B_0$, $D_0$, respectively and 
\begin{eqnarray}
2\delta M^2_{Z^0}&=&\delta A+\delta B
-{(A_0-B_0)(\delta A-\delta B)+4D_0\delta D
\over\sqrt{(A_0-B_0)^2+4D_0^2}}+2~\hat\Pi_{Z^0Z^0}(M^2_Z)\nonumber\\
&=&{M^2_W\over c^2_{(0)}}\left[-{\hat\Pi_{WW}(M_W^2)\over M^2_W}
+{s^2_{(0)}\over c^2_{(0)}}\Delta\right]+2\hat\Pi_{Z^0Z^0}(M^2_Z)
\label{eqn:oneloopMz}\\
&+&{g^2_Ee_S^2v_S^2+{g^2_Ee_H^2\over\sqrt2G_F}-{M^2_W\over c^2_{(0)}}\over
\sqrt{\dots}}
\left\{{M^2_W\over c^2_{(0)}}\left[-{\hat\Pi_{WW}(M_W^2)\over M^2_W}
+{s^2_{(0)}\over c^2_{(0)}}\Delta\right]-{g^2_Ee_H^2\over\sqrt2G_F}
\Delta_G\right\}\nonumber\\
&+&{g^2_Ee_H^2\over\sqrt2G_F}\Delta_G
-{g^2_Ee_H^2\over\sqrt{\dots}}~
{{e^2_{(0)}\over 2G_F^2s^2_{(0)}c^2_{(0)}}}
\left\{{1\over2}{s^2_{(0)}\over c^2_{(0)}}\Delta 
-{1\over2}{\hat\Pi_{WW}(M_W^2)\over M^2_W}
+{1\over2}\Delta_G\right\}\nonumber
\end{eqnarray}
where the self energies $\hat\Pi_{WW}$ and $\hat\Pi_{Z^0Z^0}$ include
the tadpole contributions. 
We would like now to demonstrate that {\it i)} in the limit 
$v_S\rightarrow\infty$ the SM result is recovered, and {\it ii)} that the 
above result is independent of the renormalization scale $\mu$.  
\vskip0.2cm

\subsection{SM limit - decoupling of the heavy $Z^\prime$ effects}

\noindent For $v_S\rightarrow\infty$ the tree level term $(M^2_{Z^0})_{(0)}$
obviously gives the SM result $M^2_W/c^2_{(0)}$. Moreover, the prefactor
in the third line of (\ref{eqn:oneloopMz}) is then $1+{\cal O}(1/v_S^4)$ and 
the prefactor of the last term is also suppressed by $1/v_S^2$. Thus in the 
limit one recovers superficially the SM formula. 
\begin{eqnarray}
2\delta M^2_{Z^0}&\rightarrow&2{M^2_W\over c^2_{(0)}}
\left[-{\hat\Pi_{WW}(M_W^2)\over M^2_W}
+{s^2_{(0)}\over c^2_{(0)}}\Delta\right]+2\hat\Pi_{Z^0Z^0}(M^2_Z)~.
\label{eqn:PiZZ-PiWW}
\end{eqnarray}
However, one still has to 
check that the appropriate combinations of $\hat\Pi_{WW}$, $\hat\Pi_{Z^0Z^0}$, 
$\Delta$ do not contain terms which would grow too fast as 
$v_S\rightarrow\infty$ invalidating the argument.

In order to show that they do not, we first note that the the $S^0$ tadpole 
${\cal T}_{S^0}$ which contributes only to $\hat\Pi_{Z^0Z^0}$ is 
suppressed (as we show below, the $h^0$ tadpoles cancel out exactly in the 
full formula (\ref{eqn:oneloopMz}) similarly as in the SM). Indeed, the $S^0$ 
coupling to $Z^0Z^0$ is proportional to $s^{\prime2}v_S\sim1/v_S^3$; the $S^0$ 
propagator is $\sim1/v_S^2$; the $S^0$ coupling to $Z^\prime Z^\prime$
and $S^0S^0$ pairs is proportional to $v_S$ so that these particles circulating
in the tadpole loop give to ${\cal T}_{S^0}$ contributions $\sim v^3_S$. 
Hence, the $S^0$ tadpole contribution to $\hat\Pi_{Z^0Z^0}$ goes as 
$\sim(1/v_S^3)(1/v_S^2)(v_S^3)\sim1/v_S^2$.

Furthermore, $\Delta$ approaches in this limit its SM form due to cancellation 
of the leading for $v_S\rightarrow\infty$ terms between $\Lambda$ and 
$\Sigma_{\nu L}+\Sigma_{eL}$ and between $\hat\Pi_{WW}(M_W^2)$ and 
$\hat\Pi_{WW}(0)$. Moreover, $\Delta_G+\hat\Pi_{WW}(M_W^2)/M^2_W$ grows only 
as $\ln(v_S^2)$, so the contribution of the last bracket in 
(\ref{eqn:oneloopMz}) vanishes for $v_S\rightarrow\infty$. Thus, in this limit 
one indeed gets (\ref{eqn:PiZZ-PiWW}) and it remains to check that the 
difference of the $Z^0$ and $W^\pm$ self-energies approaches the SM form.

For the fermionic contribution to (\ref{eqn:PiZZ-PiWW}) this is clear: for 
$\hat\Pi_{WW}(M_W^2)$ it is exactly as in the SM and that to 
$\hat\Pi_{Z^0Z^0}(M_{Z^0}^2)$ is different, but the difference is only due 
to $Z^0$ couplings which, as is follows from (\ref{eqn:Z0couplings}) and 
(\ref{eqn:s2c2}) approach as $v_S\rightarrow\infty$ their SM form. In 
particular this means that in the $SU(2)_L\times U(1)_Y\times U(1)_E$ model
the celebrated contribution $\propto m^2_t/M_W^2$ 
is present in the $M_W^2\leftrightarrow M^2_{Z^0}$ relation.

Bosonic contributions to $\hat\Pi_{WW}(M_W^2)$ and 
$\hat\Pi_{Z^0Z^0}(M_{Z^0}^2)$ individually contain terms which grow as 
$v_S\rightarrow\infty$ (the last term in the third line of 
(\ref{eqn:PIWWbosonic}) and the $Z^\prime h^0$ contribution to 
$\hat\Pi_{Z^0Z^0}$) but it is easy to check that
they cancel out in (\ref{eqn:PiZZ-PiWW}) and the difference
$\hat\Pi_{WW}(M_W^2)/M_W^2-\hat\Pi_{Z^0Z^0}(M_{Z^0}^2)/M_{Z^0}^2$
approaches its SM form too. 

Thus, we have demonstrated that in the limit $v_S\rightarrow\infty$
the finite SM expression for $M_{Z^0}$ is recovered.

\subsection{Renormalization scale $\mu$ independence of $M_{Z^0}$ at one-loop}
\noindent {\it $h^0$ tadpoles cancelation}
\vskip0.2cm

\noindent As a first step we show that the $h^0$ tadpoles ${\cal T}_{h^0}$
drop out of the formula (\ref{eqn:oneloopMz}). The contribution of 
${\cal T}_{h^0}$ to $2\hat\Pi_{Z^0Z^0}$ is
\begin{eqnarray}
2\hat\Pi_{Z^0Z^0}^{h^0~\rm tad}=2\left[{\hat e^2\over4\hat s^2\hat c^2}
-{\hat e\over\hat s\hat c}e_Hg_E~c^\prime s^\prime
-\left({\hat e^2\over4\hat s^2\hat c^2}-e_H^2g_E^2\right)s^{\prime2}\right]
\left(-2\hat v_H{{\cal T}_{h^0}\over M^2_{h^0}}\right)~.
\end{eqnarray}
With one loop accuracy and using the formulae (\ref{eqn:s2c2}) this can be
rewritten as
\begin{eqnarray}
\left[{M_W^2\over c^2_{(0)}}-
\left({M_W^2\over c^2_{(0)}}-{g^2_Ee_H^2\over\sqrt2G_F}\right)
{A_0-B_0\over\sqrt{\dots}}+ {g^2_Ee_H^2\over\sqrt2G_F}
-{e^2_{(0)}\over s^2_{(0)}c^2_{(0)}2G_F^2}{g_E^2e_H^2\over\sqrt{\dots}}\right]
\left(-{2\over\hat v_H}{{\cal T}_{h^0}\over M^2_{h^0}}\right)
\end{eqnarray}
It is then clear that each term finds in (\ref{eqn:oneloopMz}) its counterpart
with  $-\hat\Pi_{WW}^{h^0~\rm tad}/\hat M^2_W
=(2/\hat v_H)({\cal T}_{h^0}/M^2_{h^0})$ and exactly the same coefficient.
\vskip0.3cm

\noindent {\it Contribution proportional to fermion masses squared}
\vskip0.2cm

\noindent Next we consider contributions to $M^2_{Z^0}$ proportional to the 
fermion masses squared. These are hidden in $\hat\Pi_{Z^0Z^0}$,
$\hat\Pi_{WW}(M^2_W)$ and in $\hat\Pi_{WW}(0)$. As usual, they can be 
isolated by approximating the first two self energies by 
$\hat\Pi_{Z^0Z^0}(0)$ and $\hat\Pi_{WW}(0)$, respectively.
From the the formula (\ref{eqn:PIZZfermionic}) we get
\begin{eqnarray}
\Pi_{Z^0Z^0}^{\rm ferm}(q^2)
=-2\sum_fN_c^{(f)}(c^{Z^0}_{fL}-c^{Z^0}_{fR})^2~m^2_f~
b_0(0,m_f,m_f)\label{eqn:divmf2}
\end{eqnarray}
Using the couplings (\ref{eqn:Z0couplings}) and the relations 
(\ref{eqn:erels}), (\ref{eqn:s2c2}) we can write
\begin{eqnarray}
(c^{Z^0}_{fL}-c^{Z^0}_{fR})^2=
{1\over2}\left\{{\hat e^2\over4\hat s^2\hat c^2}
+\left({\hat e^2\over4\hat s^2\hat c^2}-g_E^2e_H^2\right)
{B-A\over\sqrt{\dots}}+e_H^2g_E^2
-{\hat e^2\over\hat s^2\hat c^2}{g^2_Ee^2_H\hat v_H^2\over\sqrt{\dots}}
\right\}~,\label{eqn:cfLmRsquared}
\end{eqnarray}
This makes clear that to each term in 
$2\left[\hat\Pi_{Z^0Z^0}^{\rm ferm}\right]_{\rm mass}$
there is a corresponding term with $\hat\Pi_{WW}$ in the 
formula (\ref{eqn:oneloopMz}), so that the divergences proportional
to fermion masses squared properly cancel out. Hence, the terms quadratic 
in fermion masses arising from ``oblique'' corrections are finite 
(and, hence, $\mu$-independent) just as they are in the SM. For the one-loop
top-bottom contribution using (\ref{eqn:PiWW0}) we get 
\begin{eqnarray}
M^2_{Z^0}&=&{1\over2}\left(A_0+B_0-\sqrt{(A_0-B_0)^2+4D_0^2}\right)
\nonumber\\
&-&(c^{Z^0}_{fL}-c^{Z^0}_{fR})^2{N_c\over16\pi^2}g(m_t,m_b)+
{\rm other~contributions}\phantom{aaa}
\end{eqnarray}
And in the limit $v_S\rightarrow\infty$ one recovers the SM relation
(computed using as input observables $M_W$, $G_F$ and $\alpha_{\rm EM}$).
\vskip0.3cm

\noindent {\it Remaining fermion contribution - the use of RG equations}
\vskip0.2cm

\noindent The remaining divergent fermionic contribution 
(\ref{eqn:PIZZfermionic}) to $\Pi_{Z^0Z^0}$ is proportional to $q^2$
\begin{eqnarray}
2\left[\Pi_{Z^0Z^0}^{\rm ferm}(q^2)\right]_{\rm div}^{q^2~\rm part}
={4\over3}q^2\sum_fN_c^{(f)}[(c^{Z^0}_{fL})^2+(c^{Z^0}_{fR})^2)]
\eta_{\rm div}\nonumber
\end{eqnarray}
Using the couplings (\ref{eqn:Z0couplings}) and the relations 
(\ref{eqn:erels}), (\ref{eqn:s2c2}) the right hand side takes the form
\begin{eqnarray}
&&{2\over3}M_{Z^0}^2\left\{\left(1-{A-B\over\sqrt{\dots}}\right)
{\hat e^2\over4\hat s^2\hat c^2}\left[2-4\hat s^2+8\hat s^4
+N_c(2-4\hat s^2+{40\over9}\hat s^4)\right]\right.\nonumber\\
&&\phantom{aaaaaa}+{\hat e^2\over\hat c^2}
{g^2_Ee_H\hat v^2_H\over\sqrt{\dots}}
\left[2e_l-2e_{e^c}+N_c(-{2\over3}e_q+{4\over3}e_{u^c}
-{2\over3}e_{d^c})\right]
\label{eqn:PiZZq2ferm}\\
&&\phantom{aaaaaa}+\left.\left(1+{A-B\over\sqrt{\dots}}\right)
g_E^2\left[2e_l^2+e^2_{e^c}+N_c(2e_q^2+e^2_{u^c}+e^2_{d^c})\right]
\right\}\eta_{\rm div}\nonumber
\end{eqnarray}
With one loop accuracy the prefactor of the first line can be transformed 
into 
\begin{eqnarray}
M_{Z^0}^2\left(1-{A-B\over\sqrt{\dots}}\right)=
{M_W^2\over c^2_{(0)}}+{B_0-A_0\over\sqrt{\dots}}{M_W^2\over c^2_{(0)}}
-{1\over2}{e_H^2g_E^2\over\sqrt{\dots}}{e^2_{(0)}\over s^2_{(0)}c^2_{(0)}}
{1\over2G_F^2}\nonumber
\end{eqnarray}
after which different terms arising from the first line of 
(\ref{eqn:PiZZq2ferm}) combine with the appropriate fermionic contributions to
\begin{eqnarray}
-{M^2_W\over c_{(0)}^2}\left[-{\Pi_{WW}(M_W^2)^{q^2~\rm part}\over M_W^2}
+{s_{(0)}^2\over c_{(0)}^2}\tilde\Delta\right]_{\rm div}\nonumber
\end{eqnarray}
in (\ref{eqn:oneloopMz}) canceling their divergences and 
the $\mu$ dependence exactly as in the SM.

In our renormalization scheme (outlined in Section \ref{sec:scheme}) 
the two other divergent terms in (\ref{eqn:PiZZq2ferm}) are cut off by the
$\overline{\rm MS}$ procedure. In order to see that $M^2_{Z^0}$ computed at
one loop is nevertheless renormalizations scale $\mu$ independent
we have to consider the dependence on $\mu$ of  $2(M^2_{Z^0})_{(0)}$
\begin{eqnarray}
(2M^2_{Z^0})_{(0)}=A_0+B_0(\mu)-\sqrt{\left[A_0-B_0(\mu)\right]^2
+4D_0^2(\mu)}\label{eqn:treeMz_miu}
\end{eqnarray}
The superscripts $0$ on $A$, $B$ and $D$ mean that the 
parameters $\hat e^2$, $\hat s^2$, $\hat c^2$, $\hat v_H$ have been
expressed in terms of the basic observables $\alpha_{\rm EM}$, $M_W$ and
$G_F$ to zeroth order accuracy. The $\mu$ dependence is due to
the parameters $e_Hg_E$, $e_Sg_E$, $v_S$ which are still the running 
parameters of the full theory. Using the renormalization group equations
(\ref{eqn:RGE}) and (\ref{eqn:vsRGE}) for an infinitesimal change
of scale $\mu$ we have:
\begin{eqnarray}
B_0(\mu)&=& B_0(\mu^\prime) + \delta B_1 + \delta B_2 + \delta B_v\nonumber
\end{eqnarray}
\begin{eqnarray}
4D_0^2(\mu)&=&4D_0^2(\mu^\prime)+4\delta D^2_1+4\delta D^2_2\nonumber
\end{eqnarray}
where
\begin{eqnarray}
\delta B_1&=&{1\over\sqrt2G_F}~2\left({2\over3}\sum_f e_Hg_Ee_fg_EY_H^YY_f^Y
+{1\over3}2e^2_Hg^2_EY_H^YY_H^Y\right)g^2_y~\ln{\mu^2\over\mu^{\prime2}}
\nonumber\\ 
\delta B_2&=&\left({1\over\sqrt2G_F}e^2_Hg^2_E+e_S^2g_E^2v_S^2\right)
\left({2\over3}\sum_f e_f^2g^2_E
+{1\over3}2e^2_Hg^2_E+{1\over3}e^2_Sg^2_E
\right)\ln{\mu^2\over\mu^{\prime2}}\nonumber\\ 
\delta B_v&=&e_S^2g_E^2\left(-{3\over2}\lambda_Sv_S^2+3e_S^2g_E^2v_S^2
-12{g_E^4e_S^4v_S^2+g_E^4e_S^2e_H^2v_H^2\over\lambda_S}\right)
\ln{\mu^2\over\mu^{\prime2}}\label{eqn:MZRGE}\\
4\delta D^2_1&=&{e^2_{(0)}\over s^2_{(0)}c^2_{(0)}}
{1\over2G_F^2}2\left({2\over3}\sum_f e_Hg_Ee_fg_EY_H^YY_f^Y
+{1\over3}2e^2_Hg^2_EY_H^YY_H^Y\right)g^2_y~\ln{\mu^2\over\mu^{\prime2}}
\nonumber\\ 
4\delta D^2_2&=&{e^2_{(0)}\over s^2_{(0)}c^2_{(0)}}
{1\over2G_F^2}e_H^2g_E^2\left({2\over3}\sum_f e_f^2g^2_E
+{1\over3}2e^2_Hg^2_E+{1\over3}e^2_Sg^2_E\right)
\ln{\mu^2\over\mu^{\prime2}}\nonumber
\end{eqnarray}
The formula (\ref{eqn:treeMz_miu}) then takes the form
\begin{eqnarray}
(2M^2_{Z^0})_{(0)}
&\approx&A_0+B_0(\mu^\prime)-\sqrt{\left[A_0-B_0(\mu^\prime)\right]^2
+4D_0^2(\mu^\prime)}\label{eqn:treeMz_miu_exp}
\\
&+&(\delta B_1+\delta B_2+\delta B_v)\left(1+{A_0-B_0\over\sqrt{\dots}}\right)
-{1\over2\sqrt{\dots}}\left(4\delta D^2_1+4\delta D^2_2\right)\nonumber
\end{eqnarray}
It is then a matter of some simple algebra to check that the fermion 
generation number dependent terms in (\ref{eqn:MZRGE}) precisely match the 
$\ln(1/\mu^2)$ proportional terms associated with the two last lines
of (\ref{eqn:PiZZq2ferm}) changing in these terms $\mu$ into $\mu^\prime$.
Hence, up to one loop accuracy the entire
fermionic contribution to $M^2_{Z^0}$ is renormalization scale independent.
\vskip0.3cm

\noindent {\it Renormalizations scale independence of the bosonic contribution
to $M^2_{Z^0}$}
\vskip0.2cm

\noindent The scale independence of the remaining one-loop contribution can 
be checked in a similar way (using judiciously the relations collected in
\ref{app:massmatrix}): part of the divergences with the associated $\mu$ 
dependence cancels out explicitly in the formula (\ref{eqn:oneloopMz}) as a 
result of expressing $\hat e^2$, $\hat s^2$, $\hat c^2$, $\hat v_H$ in terms 
of the basic observables $\alpha_{\rm EM}$, $M_W$ and $G_F$ with one loop 
accuracy. Other divergences are cut-off by the $\overline{\rm MS}$ 
prescription
and the explicit renormalization scale  dependence is compensated by the 
change with $\mu$ dictated by the RG of the parameters $e_kg_E$, $v_S$
in the zeroth order term $(M^2_{Z^0})_{(0)}$ (\ref{eqn:treeMz_miu}). Here
we only would like to show that the $S^0$ tadpole contribution to 
$2\hat\Pi_{Z^0Z^0}$ plays a crucial role in the working of the scheme
\cite{CHWA}. 

The couplings of $S^0$ to $S^0S^0$ and to $G^\prime G^\prime$,
$G^0G^0$ can be easily computed.\footnote{As explained in \ref{app:vsRGE}, in 
order to simplify the formulae we assume that at the scale we are 
working the scalar potential is the sum $V=V_H(H)+V_S(S)$. The physical
Higgs scalars $S^0$ and $h^0$ are then pure real parts of the singlet $S$
and of the neutral component of the doublet $H$. The $S^0$ does not couple
then to $h^0h^0$.} For the $S^0$ tadpole we then get 
\begin{eqnarray}
{\cal T}_{S^0}&=&{3\over4}\lambda_Sv_Sa(M_{S^0})
               + {1\over4}\lambda_Sv_S\tilde c^{\prime2}a(M_{G^\prime})
               + {1\over4}\lambda_Sv_S\tilde s^{\prime2}a(M_{G^0})\nonumber\\
&+&3g_E^2e_S^2 v_S \left[c^{\prime2}M^2_{Z^\prime}
\left(\ln{M^2_{Z^\prime}\over\mu^2}-{1\over3}\right)+s^{\prime2}M^2_{Z^0}
\left(\ln{M^2_{Z^0}\over\mu^2}-{1\over3}\right)\right]\nonumber
\end{eqnarray}
where $\tilde c^\prime$ and $\tilde s^\prime$ are the mixing angles 
of $G^0$ and $G^\prime$. $\tilde c^\prime$ and $\tilde s^\prime$ are 
different than $c^\prime$ and $s^\prime$ but still one has the usual 
relations $M^2_{G^0}=\xi M^2_{Z^0}$ and 
$M^2_{G^\prime}=\xi M^2_{Z^\prime}$. The $S^0$ mass is 
$M^2_{S^0}={1\over2}\lambda_Sv_S^2$. As usually we work in the Feynman
gauge $\xi=1$.

The $S^0$ tadpole gives
\begin{eqnarray}
2\hat\Pi_{Z^0Z^0}^{S^0~\rm tad}&=&2\cdot2 g_E^2e_S^2 v_S~s^{\prime2}
\left(-{{\cal T}_{S^0}\over M^2_{S^0}}\right)
=-4g_E^2e_S^2 v_S\left(1+{A-B\over\sqrt{\dots}}\right)
{1\over\lambda_Sv_S^2}
\nonumber\\
&\times&\left\{{3\over4}\lambda_Sv_S{1\over2}\lambda_Sv_S^2
+{1\over4}\lambda_Sv_Sg_E^2e_S^2v_S^2
+3g_E^2e_S^2 v_S
(g_E^2e_S^2v_S^2+g_E^2e_H^2v_H^2)\right\}\ln{1\over\mu^2}+\dots\nonumber
\end{eqnarray}
where we have used the relations
$s^{\prime2}M^2_{Z^0}+c^{\prime2}M^2_{Z^\prime}=
g_E^2e_S^2v_S^2+g_E^2e_H^2v_H^2$ and 
$\tilde s^{\prime2}M^2_{Z^0}+\tilde c^{\prime2}M^2_{Z^\prime}
=e_S^2g_E^2v_S^2$.

From $(2M_{Z^0}^2)^{\rm tree}$ (\ref{eqn:treeMz_miu}) we have instead:
\begin{eqnarray}
(2M_{Z^0}^2)^{\rm tree}&\supset&\left(1+{A_0-B_0\over\sqrt{\dots}}\right)
\delta B_v~.\nonumber
\end{eqnarray}
This explicitly shows that in the $S^0$ tadpole contribution
the scale $\mu$ is properly replaced by $\mu^\prime$ in the terms 
$\propto\lambda_S$ and 
$\propto(1/\lambda_S)$ (As we have checked, the $\lambda_S$ independent 
terms in ${\cal T}_{S^0}$ combine with the bosonic contribution 
$\hat\Pi_{Z^0Z^0}$).

We have shown, that in the one loop expression for $M^2_{Z^0}$, consistently 
with the Appelquist-Carrazzone decoupling the explicit renormalization scale
dependence is only in terms suppressed by inverse powers of $v_S$. Moreover,
the whole expression is in fact renormalization scale independent, it one
takes into account the $\mu$ dependence of the RG running of the parameters 
in the tree level term.

\section{On-shell $Z^0$ couplings to fermions}
\label{sec:Zpeak}

In this section we briefly consider the parameter $\rho$ defined in terms 
of physical $Z^0$ and $W^\pm$ masses and the Weinberg angle:
\begin{eqnarray}
\rho={M_W^2\over M^2_{Z^0}(1-\sin^2\theta^\ell_{\rm eff})}.
\label{eqn:rhoZpeakdef}
\end{eqnarray}
where $\sin^2\theta^\ell_{\rm eff}$ is defined by the form 
(\ref{eqn:onshellZ0ff}) of the effective coupling of on-shell $Z^0$ to 
fermions (we take leptons for definitness)
\begin{eqnarray}
{\cal L}_{\rm eff}^{Z^0f\bar f~{\rm on-shell}}&=&
\bar\psi_l\gamma^\mu(F_L\mathbf{P}_L+F_R\mathbf{P}_R)\psi_lZ^0_\mu~.
\label{eqn:onshellZ0ffalter}
\end{eqnarray}
Comparison of (\ref{eqn:onshellZ0ffalter}) with (\ref{eqn:onshellZ0ff}) 
gives $\sin^2\theta^\ell_{\rm eff}=F_R/2(F_R-F_L)$. 
For the formfactors $F_{L,R}$ we have the formulae
\begin{eqnarray}
F_{L,R}=-c_{\ell L,R}^{Z^0}-{1\over2}\hat\Pi_{Z^0Z^0}^\prime(M^2_{Z^0})
c_{\ell L,R}^{Z^0} 
+\hat e{\hat\Pi_{Z^0\gamma}(M^2_{Z^0})\over M^2_{Z^0}}
-{\hat\Pi_{Z^0Z^\prime}(M^2_{Z^0})\over M^2_{Z^0}-M^2_{Z^\prime}}
c_{\ell L,R}^{Z^\prime}+\dots,\label{eqn:FLFR}
\end{eqnarray}
Since we are interested only 
in the dominant universal top bottom contribution, we have not written down 
neither the genuine vertex corrections nor the final fermion self energies. 

Expressing the running coupling constants in $c_{\ell L,R}^{Z^0}$ in terms 
of $M_W^2$, $G_F$ and $\alpha_{\rm EM}$ with one loop accuracy we find
\begin{eqnarray}
c_{\ell R}^{Z^0}&=&e_{(0)}{s_{(0)}\over c_{(0)}}
\left\{1-{1\over2}\hat{\tilde\Pi}_\gamma(0)
-{\alpha_{\rm EM}\over2\pi}\ln{M^2_W\over\mu^2}
+{1\over2c^2_{(0)}}\Delta\right\}c^\prime_{(0)}\nonumber\\
&-&e_{\ell^c}g_Es^\prime_{(0)}
+e_{(0)}{s_{(0)}\over c_{(0)}}~\delta c^\prime
-e_{\ell^c}g_E~\delta s^\prime\nonumber\\
c_{\ell L}^{Z^0}&=&-{e_{(0)}\over2s_{(0)}c_{(0)}}
\left\{1-{1\over2}\hat{\tilde\Pi}_\gamma(0)
-{\alpha_{\rm EM}\over2\pi}\ln{M^2_W\over\mu^2}
+{s^2_{(0)}-c^2_{(0)}\over2c^2_{(0)}}\Delta\right\}c^\prime_{(0)}
\label{eqn:cLcRtree}\\
&+&e_{(0)}{s_{(0)}\over c_{(0)}}
\left\{1-{1\over2}\hat{\tilde\Pi}_\gamma(0)
-{\alpha_{\rm EM}\over2\pi}\ln{M^2_W\over\mu^2}
+{1\over2c^2_{(0)}}\Delta\right\}c^\prime_{(0)}\nonumber\\
&+&e_\ell g_Es^\prime_{(0)}-{e_{(0)}\over2s_{(0)}c_{(0)}}
(1-2s^2_{(0)})~\delta c^\prime+e_\ell g_E~\delta s^\prime~,\nonumber
\end{eqnarray}
where $\Delta$ is given in (\ref{eqn:Delta}). We have also introduced
$\delta c^\prime$ and $\delta s^\prime$ because original $c^\prime$ and 
$s^\prime$ depend on $\hat e$, $\hat s$, $\hat c$ and $\hat v_H^2$. The
quantities $c^\prime_{(0)}$ and $s^\prime_{(0)}$ are then given by the 
same expressions as $c^\prime$ and $s^\prime$ but with 
$\hat e$, $\hat s$, $\hat c$ and $\hat v_H^2$ replaced by 
$e_{(0)}$, $s_{(0)}$, $c_{(0)}$ and $1/\sqrt2G_F$, respectively. 

In our renormalization scheme the formfactors $F_{L,R}$ given by 
(\ref{eqn:FLFR}) and (\ref{eqn:cLcRtree}) are finite if the 
$\overline{\rm MS}$ scheme is employed.  Moreover
their nonvanishing as $v_S\rightarrow\infty$ parts are renormalization 
scale independent (i.e. they are just finite) and the explicit $\mu$ 
dependence of the one-loop terms is compensated by the change of 
the running parameters $e_Hg_E$, $e_\ell g_E$, $e_{\ell^c}g_E$ and $v_S$ 
entering the zeroth order contributions. 

For $\delta c^\prime$ and $\delta s^\prime$ we find
\begin{eqnarray}
\delta s^\prime=-{c^\prime_{(0)}\over4s^\prime_{(0)}}\delta c^\prime
&=&{1\over4s^\prime_{(0)}}
{1\over(\sqrt{\dots})^3}\left[4D_0^2(\delta A-\delta B)-
(A_0-B_0)4D_0\delta D\right]\nonumber\\
&=&{c^\prime_{(0)}\over(\sqrt{\dots})^2}
{e_{(0)}\over2s_{(0)}c_{(0)}}{e_He_S^2g^3_Ev_S^2\over\sqrt2G_F}
\left[-{\hat\Pi_{WW}(0)\over M^2_W}\right]+\dots
\end{eqnarray}
where in the second line, in order to isolate the dominant top-bottom 
contributions to the formfactors $F_L$ and $F_R$, we have isolated only the 
term with $\hat\Pi_{WW}(0)$. Combining this with 
\begin{eqnarray}
\hat\Pi_{Z^0Z^\prime}(M^2_{Z^0})\approx
\hat\Pi_{Z^0Z^\prime}(0)=-\sum_f
(c_{fL}^{Z^0}-c_{fR}^{Z^0})(c_{fL}^{Z^\prime}-c_{fR}^{Z^\prime})
2N_c^{(f)}m_f^2b_0(0,m_f,m_f)\nonumber\\
=-{1\over(\sqrt{\dots})}{\hat e\over2\hat s\hat c}e_He_S^2g_E^3v^2_S
\sum_f2N_c^{(f)}m_f^2b_0(0,m_f,m_f)
\end{eqnarray}
(where again we have used the results (\ref{eqn:Z0couplings}),
(\ref{eqn:ZPcouplings}) and (\ref{eqn:usefulrel1})) and using the fact that
$M^2_{Z^0}-M^2_{Z^\prime}=-\sqrt{\dots}$ we find
\begin{eqnarray}
F_{L,R}^{t,b}\approx-{1\over(\sqrt{\dots})^2}
{\hat e\over2\hat s\hat c}e_He_S^2g_E^3v^2_S c^{Z^\prime}_{\ell L,R}
{N_c\over16\pi^2}g(m_t,m_b)~.
\end{eqnarray}
Since $(\sqrt{\dots})^2\equiv(A_0-B_0)^2+4D_0^2\sim v^4_S$ as
$v_S\rightarrow\infty$, this contribution is explicitly suppressed in 
this limit. It is easy to see that the expressions for $F_L$ and $F_R$ 
(\ref{eqn:FLFR}), (\ref{eqn:cLcRtree}) do not involve any other contributions 
proportional to $m^2_t$ and $m_b^2$ and, therefore, no contribution
$\propto m_t^2/M_W^2$ enter $\sin^2\theta_{\rm eff}^\ell$ at one 
loop.\footnote{In the SM the formfactors $F_L$ and $F_R$ do not receive
any such contribution if the scheme based on $M_W$, $G_F$ and
$\alpha_{\rm EM}$ as input observables is employed.}
Since we have already shown that for $v_S\rightarrow\infty$ one 
recovers also the SM expression for $M_{Z^0}$, we conclude, that 
in the $U(1)_Y\times U(1)_E$ model 
\begin{eqnarray}
\rho={M_W^2\over M^2_{Z^0}(1-\sin^2\theta^\ell_{\rm eff})}
=1 + {N_c\over16\pi^2}\sqrt2G_Fg(m_t,m_b)+{\cal O}(m^2_t/v_S^2)+\dots
\end{eqnarray}
where dots stand for other SM contribution as well as for other terms 
suppressed in the limit $v_S\rightarrow\infty$ (also those arising from the 
tree level contribution contributon to $\rho$ (see eq. (\ref{eqn:rhoupto})).
Similar result can be proven also for $\rho_{Zf}$ defined by 
the effective Lagrangian (\ref{eqn:onshellZ0ff}).

It should be stressed that unlike $\rho_{\rm low}$ to which one loop 
corrections have been computed  in 
section 5, the parameter $\rho$ defined in (\ref{eqn:rhoZpeakdef})
is not equal to unity at the tree level. Therefore the one loop result
for $\rho$ does depend on the renormalization scheme and in particular on
the chosen set of input observables. 
This observation is helpful in understanding the apparent discrepancy of
our results with the claim of refs. \cite{JEG,CZGLJEZR,DAWSON} that in models 
like the one considered here the contribution to $\rho$ proportonal to 
$m_t^2/M^2_W$ is absent. Refs. \cite{JEG,CZGLJEZR,DAWSON} use 
$\sin^2\theta^\ell_{\rm eff}$ as one of the input observables and then, as 
we have, commented earlier, explicit Appelquist-Carrazzone decoupling is lost.
However, our point is that the renormalization 
scheme can be chosen in such a way that new physics effects can be treated 
as corrections to the well established SM resuls.

\section{Conclusions}

In this paper we have discussed some technical aspects related to the 
$U(1)_E$ extension of the standard electroweak theory. We have elucidated 
the correct treatement of the additional coupling constants and presented 
the one loop renormalization group equations in the form adapted to 
practical calculations. Furthermore we have proposed a renormalization 
scheme employing as in the SM only three input observables (for technical 
convenience we have chosen to work with $M_W$, $G_F$ and $\alpha_{\rm EM}$ 
instead of the customary set  $M_{Z^0}$, $G_F$ and $\alpha_{\rm EM}$) 
which has the virtue of making the decoupling of havy 
$Z^\prime$ effects manifest. To demonstrate this we have 
computed the parameter $\rho$ defined either in terms of the low energy
neutrino scattering processes or in terms of physical $M^2_W$, $M^2_{Z^0}$
and $\sin^2\theta^l_{\rm eff}$ as measured in $Z^0\rightarrow l^+l^-$. In 
addition, in both cases we have shown explicitly in a renormalization scheme 
in which the Appelquist-Carrazzone decoupling is manifest
the $\propto G_Fm^2_t$ contribution to the $\rho$ parameters is
present and up to terms vanishing as $M_{Z^\prime}\rightarrow\infty$
take the form as in the SM. Our calculation supports therefore similar
observation made in \cite{AGMAPEVI} long time ago.

Our choice of $M_W$, $G_F$ and $\alpha_{\rm EM}$ as input observables
instead of the commonly used set $M_Z$, $G_F$ and $\alpha_{\rm EM}$
was dictated by the desire of demonstrating crucial aspects of our 
renormalization scheme (in particular the role of the renormalization
group equations in proving scale independence of computed observables)
analytically. We have checked however, that the explicit decoupling of 
heavy $Z^\prime$ effects (that the
expressions for the electroweak observables approach their SM form for
$v_S\propto M_{Z^\prime}\rightarrow\infty$), do not depend on whether one 
uses $M_W$ or $M_Z$.

The Appelquist-Carrazzone decoupling offers a possibility of a systematic 
inclusion of all large logarithmic $\sim[\ln(M_{Z^\prime}/M_{Z^0})]^n$ 
corrections by taking into account the RG running of the Wilson coeffcients
of nonrenormalizable operators generated by decoupling of the heavy 
$Z^\prime$ sector.

\vskip1.0cm
\section*{Acknowledgments}

P.H.Ch. would like to thank the CERN Theory Group for hospitality.
P.H.Ch., and S.P. were partially supported by the European Community 
Contract MRTN-CT-2004-503369 for years 2004--2008 and by the Polish KBN 
grant 1 P03B 099 29 for years 2005--2007. 

\newpage
\renewcommand{\thesection}{Appendix~\Alph{section}}
\renewcommand{\theequation}{\Alph{section}.\arabic{equation}}
\setcounter{section}{0}

\section{Useful formulae}
\label{app:massmatrix}
\setcounter{equation}{0}

The mass matrix of $Z^0$ and $Z^\prime$ which arises as a $2\times2$ submatrix
after rotating (\ref{eqn:neutralgbmassmatrix}) by the angle $\theta_W$ reads
\begin{eqnarray}
\left(\matrix{
{1\over4}(g_y^2+g_2^2)v_H^2 &-{1\over2}\sqrt{g_y^2+g_2^2}g_Ee_Hv_H^2\cr
-{1\over2}\sqrt{g_y^2+g_2^2}g_Ee_Hv_H^2&e_H^2g_E^2v_H^2+e_S^2g_E^2v_S^2}
\right)
=\left(\matrix{A&D\cr D&B}\right)\nonumber
\end{eqnarray}
It is diagonalized by the rotation by the angle $\theta^\prime$ determined 
from (\ref{eqn:mixingdef}). For $s^{\prime2}$, $c^{\prime2}$ and 
$s^\prime c^\prime$ one derives the following usefull expressions
\begin{eqnarray}
s^{\prime2}&=&{1\over2}\left(1+{A-B\over\sqrt{(A-B)^2+4D^2}}\right)
={1\over4}{g_y^2+g_2^2\over g^2_E}{e^2_Hv^4_H\over e_S^4v_S^4}+\dots~,
\nonumber\\
c^{\prime2}&=&{1\over2}\left(1-{A-B\over\sqrt{(A-B)^2+4D^2}}\right)
=1-{1\over8}{g_y^2+g_2^2\over g^2_E}{e^2_Hv^4_H\over e_S^4v_S^4}+\dots~,
\label{eqn:s2c2}\\
s^\prime c^\prime &=&{-D\over\sqrt{(A-B)^2+4D^2}}
={1\over2}{\sqrt{g_y^2+g_2^2}\over g_E}{e_Hv^2_H\over e_S^2v_S^2}+
\dots\nonumber
\end{eqnarray}
Other useful expressions are
\begin{eqnarray}
&&{s^{\prime2}\over M^2_{Z^0}}+
  {c^{\prime2}\over M^2_{Z^\prime}}
={1\over M^2_{Z^0}M^2_{Z^\prime}}
\left({e^2\over4s^2c^2}v^2_H\right)\nonumber\\
&&{c^{\prime2}\over M^2_{Z^0}}+
  {s^{\prime2}\over M^2_{Z^\prime}}
={1\over M^2_{Z^0}M^2_{Z^\prime}}
\left(g^2_Ee_H^2v_H^2+g^2_Ee_S^2v_S^2\right)\label{eqn:usefulrel1}\\
&&s^\prime c^\prime\left({1\over M^2_{Z^0}}-{1\over M^2_{Z^\prime}}\right)
={1\over M^2_{Z^0}M^2_{Z^\prime}}\left({e\over2sc}
g_Ee_Hv^2_H\right)
\nonumber
\end{eqnarray}
and 
\begin{eqnarray}
M^2_{Z^0}M^2_{Z^\prime}={e^2\over4s^2c^2}g_E^2e_S^2v^2_Hv^2_S
\label{eqn:usefulrel2}
\end{eqnarray}
Other useful relations are
\begin{eqnarray}
c^{\prime2}M^2_{Z^0}=Ac^{\prime2}+Ds^\prime c^\prime\nonumber\\
s^{\prime2}M^2_{Z^0}=Bs^{\prime2}+Ds^\prime c^\prime\nonumber\\
c^{\prime2}M^2_{Z^\prime}=Bc^{\prime2}-Ds^\prime c^\prime\\
s^{\prime2}M^2_{Z^\prime}=As^{\prime2}-Ds^\prime c^\prime\nonumber
\end{eqnarray}

\section{Calculation of the input observables $\alpha_{\rm EM}$, $G_F$ and 
$M_W$}
\label{app:input}
\setcounter{equation}{0}

Here we outline the calculation in the $SU(2)_L\times U(1)_Y\times U(1)_E$
model of the basic input observables $\alpha_{\rm EM}$, $G_F$ and $M_W$. 
The formula for $M_W$ is simple
\begin{eqnarray}
M_W^2={\hat e^2\over4\hat s^2}\hat v_H^2+\hat\Pi_{WW}(M_W^2)
\end{eqnarray}
where $\hat\Pi_{WW}(M_W^2)$ includes in principle also the tadpole 
contribution. Expressions for $\alpha_{\rm EM}$ and  $G_F$ are derived below.

\subsection{Calculation of $\delta\alpha_{\rm EM}$}
\label{app:alphaEM}

This is most easily computed using the effective Lagrangian technique 
\cite{POK}. Below the electroweak scale (the renormalizable part of) the 
effective Lagrangian for electromagnetic interactions has the form
\begin{eqnarray}
&&{\cal L}=-{1\over4}(1+\delta z_\gamma) f_{\mu\nu}f^{\mu\nu}\nonumber\\
&&\phantom{aaa}
+(1+\delta z_2^L)\bar\psi_ei\not\!\partial~\mathbf{P}_L\psi_e
-(e+\delta e + \hat e ~\delta z_2^L+{1\over2}\hat e\delta z_\gamma)
\bar\psi_e~q_e\not\!\! A~\mathbf{P}_L\psi_e\label{eqn:effLagr}\\
&&\phantom{aaa}
+(1+\delta z_2^R)\bar\psi_ei\not\!\partial~\mathbf{P}_R\psi_e
-(\hat e+\delta e + \hat e ~\delta z_2^R+{1\over2}\hat e~\delta z_\gamma)
\bar\psi_e~q_e\not\!\! A~\mathbf{P}_R\psi_e\nonumber\\
&&\phantom{aaa}+{\rm counterterms}~.\nonumber
\end{eqnarray}
$\hat e + \delta e$ is the electromagnetic coupling of
QED at the scale just below the Fermi scale threshold; it can be easily
related to $\alpha_{\rm EM}$ via the RG running.

The factors $\delta z_2^L$ and $\delta z_2^R$ are such that they reproduce 
at the tree level contributions of virtual  $W$, $Z^0$ and $Z^\prime$ 
to the electron self-energies (computed at zero momentum). Similarly,
\begin{eqnarray}
\delta z_\gamma = -[\tilde\Pi_\gamma(0)]_{W,G^+,f}\label{eqn:zgamm_def}
\end{eqnarray}
reproduces at the tree level the vacuum polarization due to decoupled 
heavy particles $W^\pm$ and top quark.

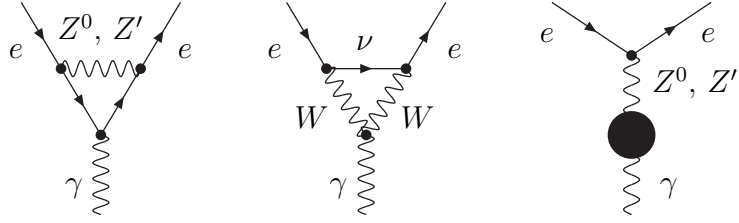
\begin{figure}[htbp]
\begin{center}
\begin{picture}(300,120)(0,0)
\ArrowLine(10,90)(25,65)
\ArrowLine(25,65)(40,40)
\ArrowLine(40,40)(55,65)
\ArrowLine(55,65)(70,90)
\Vertex(25,65){2}
\Vertex(55,65){2}
\Vertex(40,40){2}
\Photon(25,65)(55,65){3}{5}
\Photon(40,40)(40,10){3}{5}
\Text(08,72)[]{$e$}
\Text(72,72)[]{$e$}
\Text(40,80)[]{$Z^0$, $Z^\prime$}
\Text(30,20)[]{$\gamma$}
\ArrowLine(110,90)(125,65)
\ArrowLine(125,65)(155,65)
\ArrowLine(155,65)(170,90)
\Vertex(125,65){2}
\Vertex(155,65){2}
\Vertex(140,40){2}
\Photon(125,65)(140,40){3}{5}
\Photon(140,40)(155,65){3}{5}
\Photon(140,40)(140,10){3}{5}
\Text(106,72)[]{$e$}
\Text(174,72)[]{$e$}
\Text(140,75)[]{$\nu$}
\Text(120,47)[]{$W$}
\Text(160,47)[]{$W$}
\Text(130,20)[]{$\gamma$}
\ArrowLine(210,90)(240,70)
\ArrowLine(240,70)(270,90)
\Vertex(240,70){2}
\Photon(240,70)(240,10){3}{8}
\Vertex(240,40){9}
\Text(209,78)[]{$e$}
\Text(270,78)[]{$e$}
\Text(265,60)[]{$Z^0$, $Z^\prime$}
\Text(255,20)[]{$\gamma$}
\end{picture}
\end{center}
\caption{Corrections to the photon-electron vertex in a model with extra 
$U(1)$. The external line momenta can be off-shell but must be $\ll M_Z$.}
\label{fig:photonvertexcorrinextrau1}
\end{figure}

The vertex corrections determining the combinations
$\delta e+\hat e~\delta z_2^{L,R}+{1\over2}\hat e~\delta z_\gamma$ are 
shown in figure \ref{fig:photonvertexcorrinextrau1}. Owing to the $U(1)_Y$ 
and $U(1)_E$ Ward identities the $Z^0$ and $Z^\prime$ contributions to 
$\delta e$ are exactly canceled by the $Z^0$ and $Z^\prime$ contributions 
to $\delta z_2^L$ and $\delta z_2^R$, respectively. The second diagram in  
figure \ref{fig:photonvertexcorrinextrau1} is exactly as in the SM and
combines with the $W$ contribution to $\delta z_2^L$. As a result 
from the photon coupling to left-chiral electrons one gets 
\begin{eqnarray}
&&\delta e={1\over2}\hat e~\tilde\Pi_\gamma(0)
+c^{Z^0}_{eL}{\hat\Pi_{\gamma Z^0}(0)\over M^2_{Z^0}}
+c^{Z^\prime}_{eL}{\hat\Pi_{\gamma Z^\prime}(0)\over M^2_{Z^\prime}}
-{\hat e^3\over16\pi^2\hat s^2}
\left(\eta_{\rm div}+\ln{\hat M^2_W\over\mu^2}\right)\nonumber
\end{eqnarray}
The self energies $\hat\Pi_{\gamma Z^0}(0)$ and 
$\hat\Pi_{\gamma Z^\prime}(0)$ receive contributions only from the vitrual
$W^+W^-$ and $W^\pm G^\mp$ pairs. We get
\begin{eqnarray}
&&\delta e={1\over2}\hat e~\tilde\Pi_\gamma(0)
-{1\over16\pi^2}{\hat e^3\over\hat s^2}\left(\eta_{\rm div}+
\ln{\hat M^2_W\over\mu^2}\right)\nonumber\\
&&\phantom{aaa}
+{1\over16\pi^2}2c^{Z^0}_{eL}
\left[-\hat e^2{\hat c\over\hat s}c^\prime
-\hat e\left(\hat e{\hat s\over\hat c}c^\prime-2e_Hg_Es^\prime\right)\right]
{\hat M^2_W\over M^2_{Z^0}}\ln{\hat M^2_W\over\mu^2}\nonumber\\
&&\phantom{aaa}
+{1\over16\pi^2}2c^{Z^\prime}_{eL}
\left[\hat e^2{\hat c\over\hat s}s^\prime
+\hat e\left(\hat e{\hat s\over\hat c}s^\prime+2e_Hg_Ec^\prime\right)\right]
{\hat M^2_W\over M^2_{Z^\prime}}\ln{\hat M^2_W\over\mu^2}
\nonumber
\end{eqnarray}
By using the relation (\ref{eqn:usefulrel1}), (\ref{eqn:usefulrel2}) this 
can be reduced to
\begin{eqnarray}
\delta e ={1\over2}\hat e~\tilde\Pi_\gamma(0)
-{\hat e^3\over8\pi^2}\left(\eta_{\rm div}+\ln{\hat M^2_W\over\mu^2}\right)
\nonumber
\end{eqnarray}
which (as could be expected) is the same as in the SM. The same result
is obtained by considering the photon coupling to right-chiral electron.

\subsection{Calculation of $\delta G_F$}

Calculation of $\delta G_F$ proceeds as in the SM. The only modification 
is that there are additional box diagrams with $Z^\prime$ and in addition 
the $W$ boson self energy $\Pi_{WW}(q^2)$ as well as the self energies of 
external line fermions are modified by the presence of $Z^\prime$ (there 
are contributions from virtual $Z^\prime$ and the couplings of $Z^0$ are 
modified). Still the formula takes the form
\begin{eqnarray}
G_F={1\over\sqrt2\hat v^2_H}(1+\Delta_G)=
{\hat e^2\over4\sqrt2\hat s^2\hat M_W^2}(1+\Delta_G)
\nonumber
\end{eqnarray}
with $\Delta_G$ given by (\ref{eqn:DeltaG})
\begin{eqnarray}
\Delta_G = 
-{\hat\Pi_{WW}(0)\over\hat M^2_W}+B_{W\gamma}+B_{WZ^0}+B_{WZ^\prime}
+2\hat\Lambda+\hat\Sigma_{eL}+\hat\Sigma_{\nu L}~.\label{eqn:DeltaG}
\end{eqnarray}
Here $B_{W\gamma}$ is the contribution (in units of the tree level
$W$ exchange) of the $W\gamma$ box with subtracted 
contribution of the photonic vertex correction to thetree level
diagram in the low energy effective four-Fermi theory of $\mu^-$ decay
\begin{eqnarray}
B_{W\gamma}={\hat e^2\over16\pi^2}\left(\eta_{\rm div}+
{1\over2}+\ln{M^2_W\over\mu^2}\right)
\nonumber
\end{eqnarray}
(this contribution is the same as in the SM) and $B_{WZ^0}$ and 
$B_{WZ^\prime}$ denote the contributions of the box diagrams with $WZ^0$ and 
$WZ^\prime$, respectively: 
\begin{eqnarray}
B_{WZ^0}={1\over16\pi^2}\left[\left(c^{Z^0}_{eL}\right)^2
+\left(c^{Z^0}_{\nu L}\right)^2-8~c^{Z^0}_{eL}c^{Z^0}_{\nu L}\right]
{M^2_W\over M^2_W-M^2_{Z^0}}\ln{M^2_W\over M^2_{Z^0}}\nonumber
\end{eqnarray}
and $B_{WZ^\prime}$ is given by a similar expression with 
$c^{Z^0}_{e,\nu L}\rightarrow c^{Z^\prime}_{e,\nu L}$ and 
$M^2_{Z^0}\rightarrow M^2_{Z^\prime}$.

For the contributions $\hat\Lambda^{(i)}$ of individual diagrams to the vertex 
corrections $\hat\Lambda=(1/16\pi^2)\sum_i\hat\Lambda^{(i)}$ one finds:
\begin{eqnarray}
&&\hat\Lambda^{Z^0e\nu}
=-c^{Z^0}_{eL}c^{Z^0}_{\nu L}
\left(\eta_{\rm div}+{1\over2}+\ln{\hat M^2_{Z^0}\over\mu^2}\right)\nonumber\\
&&\hat\Lambda^{Z^\prime e\nu}
=-c^{Z^\prime}_{eL}c^{Z^\prime}_{\nu L}
\left(\eta_{\rm div}+{1\over2}+\ln{\hat M^2_{Z^\prime}\over\mu^2}\right)
\nonumber\\
&&\hat\Lambda^{\nu WZ^0}
=-3c^{Z^0}_{\nu L}
\left(\hat e{\hat c\over\hat s}c^\prime\right)
\left(\eta_{\rm div}-{5\over6}+\ln{\hat M^2_W\over\mu^2}
+{\hat M^2_{Z^0}\over \hat M^2_{Z^0}-\hat M^2_W}
\ln{\hat M^2_{Z^0}\over\hat M^2_W}\right)\nonumber\\
&&\hat\Lambda^{\nu WZ^\prime}
=-3c^{Z^\prime}_{\nu L}
\left(-\hat e{\hat c\over\hat s}s^\prime\right)
\left(\eta_{\rm div}-{5\over6}+\ln{\hat M^2_W\over\mu^2}
+{\hat M^2_{Z^\prime}\over \hat M^2_{Z^\prime}-\hat M^2_W}
\ln{\hat M^2_{Z^\prime}\over\hat M^2_W}\right)\nonumber\\
&&\hat\Lambda^{eZ^0W}
=3c^{Z^0}_{eL}\left(\hat e{\hat c\over\hat s}c^\prime\right)
\left(\eta_{\rm div}-{5\over6}+\ln{\hat M^2_W\over\mu^2}
+{\hat M^2_{Z^0}\over \hat M^2_{Z^0}-\hat M^2_W}
\ln{\hat M^2_{Z^0}\over\hat M^2_W}\right)\nonumber\\
&&\hat\Lambda^{eZ^\prime W}
=3c^{Z^\prime}_{eL}\left(-\hat e{\hat c\over\hat s}s^\prime\right)
\left(\eta_{\rm div}-{5\over6}+\ln{\hat M^2_W\over\mu^2}
+{\hat M^2_{Z^\prime}\over \hat M^2_{Z^\prime}-\hat M^2_W}
\ln{\hat M^2_{Z^\prime}\over\hat M^2_W}\right)\nonumber\\
&&\hat\Lambda^{e\gamma W}
=-3\hat e^2
\left(\eta_{\rm div}-{5\over6}+\ln{\hat M^2_W\over\mu^2}\right)\nonumber
\end{eqnarray}
so that the divergent part of $\hat\Lambda$ is
\begin{eqnarray}
\hat\Lambda_{\rm div}={1\over16\pi^2}\left(\hat e^2
{1-2\hat s^2-12c^2\over4\hat s^2\hat c^2}-e_l^2g_E^2\right)\eta_{\rm div}
\nonumber
\end{eqnarray}

Finally, for the self energies $\hat\Sigma_{\nu L}$ and 
$\hat\Sigma_{eL}$ of the left-chiral electron 
and neutrino, respectively  one gets
\begin{eqnarray}
&&16\pi^2\hat\Sigma_{\nu L}={\hat e^2\over2\hat s^2}
\left(\eta_{\rm div}+{1\over2}+\ln{\hat M^2_W\over\mu^2}\right)
+\left(c^{Z^0}_{\nu L}\right)^2
\left(\eta_{\rm div}+{1\over2}+\ln{\hat M^2_{Z^0}\over\mu^2}\right)
\nonumber\\
&&\phantom{aaaaaaaaaaaaaaaaaaaaa}
+\left(c^{Z^\prime}_{\nu L}\right)^2
\left(\eta_{\rm div}+{1\over2}+\ln{\hat M^2_{Z^\prime}\over\mu^2}\right)
\nonumber\\
&&16\pi^2\hat\Sigma_{e L} = {\hat e^2\over2\hat s^2}
\left(\eta_{\rm div}+{1\over2}+\ln{\hat M^2_W\over\mu^2}\right)
+\left(c^{Z^0}_{e L}\right)^2
\left(\eta_{\rm div}+{1\over2}+\ln{\hat M^2_{Z^0}\over\mu^2}\right)
\nonumber\\
&&\phantom{aaaaaaaaaaaaaaaaaaaaa}
+\left(c^{Z^\prime}_{e L}\right)^2
\left(\eta_{\rm div}+{1\over2}+\ln{\hat M^2_{Z^\prime}\over\mu^2}\right)
\nonumber
\end{eqnarray}
with the divergent part 
\begin{eqnarray}
(\hat\Sigma_{\nu L}+\hat\Sigma_{e L})_{\rm div}={1\over16\pi^2}\left(
{\hat e^2\over\hat s^2}+{\hat e^2\over4\hat s^2\hat c^2}
[1 + (1-2\hat s^2)^2] + 2e_l^2g_E^2\right)\eta_{\rm div}\nonumber
\end{eqnarray}

Collecting all divergent parts together we get for boxes, vertex and self 
energy corrections exactly the same divergent part as in the SM
\begin{eqnarray}
(B_{\rm boxes}+2\hat\Lambda+\hat\Sigma_{e L}
+\hat\Sigma_{\nu L})_{\rm div}=
-{\hat e^2\over16\pi^2}{4\over\hat s^2}\eta_{\rm div}
\label{eqn:boxesandvertices}
\end{eqnarray}

\section{RG equation for $v_S$}
\label{app:vsRGE}
\setcounter{equation}{0}

The most general scalar field potential in the model considered in this
paper is
\begin{eqnarray}
V=m^2_S S^\star S+{\lambda_S\over4}(S^\star S)^2
+m^2_H H^\dagger H+{\lambda_H\over4}(H^\dagger H)^2
+\kappa(S^\star S)(H^\dagger H)\nonumber
\end{eqnarray}
In order to simplify the formulae we have assumed that at one particular
renormalization scale $\mu$, at which we chose to work, $\kappa(\mu)=0$. 
However, to derive the renormalization group equation for $v_S$ one has to 
keep $\kappa$. With 
\begin{eqnarray}
S={1\over\sqrt2}(v_S+S^0+iG_S)\phantom{aaaaaa}
H={1\over\sqrt2}\left(\matrix{\sqrt2G^+\cr v_H+h^0+iG_H}\right)
\label{eqn:HandSdecomposition}
\end{eqnarray}
(where $h^0$ and $S^0$ are the physical Higgs scalars and $G_H$ and $G_S$ are 
the fields whose appropriate linear combinations $G^0$ and $G^\prime$
become the longitudinal components of the massive $Z^0$ and 
$Z^\prime$), the formulae determining $v_S^2$ and $v_H^2$ read
\begin{eqnarray}
m_H^2+{1\over4}\lambda_Hv^2_H+{1\over2}\kappa v_S^2=0\nonumber\\
m_S^2+{1\over4}\lambda_Sv^2_S+{1\over2}\kappa v_H^2=0\label{eqn:vSvHgeneral}
\end{eqnarray}
Differentiating the second one with respect to $\mu$ we get at $\kappa=0$:
\begin{eqnarray}
\mu{dv_S^2\over d\mu}=-{4\over\lambda_S}\left(\mu{dm_S^2\over d\mu}+
{1\over4}v^2_S~\mu{d\lambda_S\over d\mu}+{1\over2}v_H^2~\mu{d\kappa\over d\mu}
\right)\label{eqn:dvs2overdt}
\end{eqnarray}
Thus, to find the derivative of $v_S^2$ at the scale $\mu$ such, that
$\kappa(\mu)=0$ we need to get also $d\kappa/dt$. Calculating derivatives
appering in the right hand side of (\ref{eqn:dvs2overdt}) is standard
\begin{eqnarray}
\mu{d\over d\mu}\lambda_S&=&2\epsilon\lambda_S
+5\lambda_S^2-12\lambda_Sg_E^2e_S^2+24g_E^4e_S^4\nonumber\\
\mu{d\over d\mu}m^2_S&=&m^2_S(2\lambda_S-6g_E^2e_S^2)\\
\mu{d\over d\mu}\kappa&=&12g_E^4e_S^2e_H^2\nonumber
\end{eqnarray}

Using these results and (\ref{eqn:dvs2overdt}) it is easy to derive
\begin{eqnarray}
\mu{d\over d\mu} v_S^2=v_S^2\left(-3\lambda_S+6g_E^2e_S^2\right)
-24{g_E^4e_S^4v_S^2+g_E^4e_S^2e_H^2v_H^2\over\lambda_S}\label{eqn:vsRGE}
\end{eqnarray}

\section{Vector boson self energies}
\label{app:VVself}

The fermionic one loop contribution to $\Pi_{WW}(q^2)$ in the 
$SU(2)\times U(1)_E\times U(1)_Y$
is as in the SM. For the bosonic part of $\Pi_{WW}(q^2)$ we have
\begin{eqnarray}
&&-{\hat e^2\over\hat s^2}\tilde A(q^2,\hat M_W,\hat M_{h^0})
-{\hat e^2\over\hat s^2}\tilde A(q^2,\hat M_W,\hat M_{Z^0})\nonumber\\
&&+{\hat e^2\over\hat s^2}\hat M^2_Wb_0(q^2,\hat M_W,\hat M_{h^0})
+\hat e^2\hat M^2_Wb_0(q^2,\hat M_W,0)\nonumber\\
&&+\left(-\hat e{\hat s\over\hat c}c^\prime+2e_Hg_Es^\prime\right)^2
\hat M^2_Wb_0(q^2,\hat M_W,\hat M_{Z^0})
+\left(\hat e{\hat s\over\hat c}s^\prime+2e_Hg_Ec^\prime\right)^2
\hat M^2_Wb_0(q^2,\hat M_W,\hat M_{Z^\prime})\nonumber\\
&&-\hat e^2{\hat c^2\over\hat s^2}c^{\prime^2}
\left[8\tilde A(q^2,\hat M_W,\hat M_{Z^0})
+(4q^2+\hat M^2_W+\hat M^2_{Z^0})b_0(q^2,\hat M_W,\hat M_{Z^0})
-{2\over3}{q^2\over16\pi^2}\right]\nonumber\\
&&-\hat e^2{\hat c^2\over\hat s^2}s^{\prime^2}
\left[8\tilde A(q^2,\hat M_W,\hat M_{Z^\prime})
+(4q^2+\hat M^2_W+\hat M^2_{Z^\prime})b_0(q^2,\hat M_W,\hat M_{Z^\prime})
-{2\over3}{q^2\over16\pi^2}\right]\nonumber\\
&&-\hat e^2\left[8\tilde A(q^2,\hat M_W,0)
+(4q^2+\hat M^2_W)b_0(q^2,\hat M_W,0)
-{2\over3}q^2{q^2\over16\pi^2}\right]\label{eqn:PIWWbosonic}
\end{eqnarray}
The divergent part of this contribution taken at $q^2=0$ is
\begin{eqnarray}
16\pi^2[\hat\Pi_{WW}(0)]^{\rm bos}_{\rm div}
=\left(\hat e^2{\hat s^2-\hat c^2\over\hat s^2\hat c^2}\hat M^2_W
+4e_H^2g_E^2\hat M^2_W\right)\eta_{\rm div}\label{eqn:WWat0div}
\end{eqnarray}
(we have used
$c^{\prime^2}\hat M^2_{Z^0}+s^{\prime^2}\hat M^2_{Z^\prime}=
\hat M^2_W/\hat c^2$). It differs from the SM only by the last term.

Below we list all bosonic contributions to $\Pi_{Z_1Z_2}(q^2)$ for 
$Z_1Z_2=Z^0Z^0$, $Z^\prime Z^\prime$, $Z^0Z^\prime$ 
\begin{eqnarray}
W^+W^-:\phantom{aa}
-\hat e^2{\hat c^2\over\hat s^2}\left[8\tilde A(q^2,\hat M_W,\hat M_W)
+(4q^2+2\hat M^2_W)b_0(q^2,\hat M_W,\hat M_W)-{2\over3}{q^2\over16\pi^2}
\right]\times
\left(\matrix{c^{\prime2}\cr s^{\prime2}\cr-c^\prime s^\prime}\right)
\nonumber
\end{eqnarray}
\begin{eqnarray}
G^\pm W^\mp:\phantom{aaaaa}
+2\hat M^2_Wb_0(q^2,\hat M_W,\hat M_W)\times\left(\matrix{
(-\hat e{\hat s\over\hat c}c^\prime+2e_Hg_Es^\prime)^2\cr
(\hat e{\hat s\over\hat c}s^\prime+2e_Hg_Ec^\prime)^2\cr
(-\hat e{\hat s\over\hat c}c^\prime+2e_Hg_Es^\prime)
(\hat e{\hat s\over\hat c}s^\prime+2e_Hg_Ec^\prime)}\right)\nonumber
\end{eqnarray}
\begin{eqnarray}
G^+G^-:\phantom{aaaaa}
-\tilde A(q^2,\hat M_W,\hat M_W)\times\left(\matrix{
(\hat e{\hat c^2-\hat s^2\over\hat s\hat c}c^\prime+2e_Hg_Es^\prime)^2\cr
(-\hat e{\hat c^2-\hat s^2\over\hat s\hat c}s^\prime+2e_Hg_Ec^\prime)^2\cr
(\hat e{\hat c^2-\hat s^2\over\hat s\hat c}c^\prime+2e_Hg_Es^\prime)
(-\hat e{\hat c^2-\hat s^2\over\hat s\hat c}s^\prime+2e_Hg_Ec^\prime)}\right)
\nonumber
\end{eqnarray}
\begin{eqnarray}
G^0h^0:\phantom{aaaaa}
-\tilde A(q^2,\hat M_{Z^0},\hat M_{h^0})\times\left(\matrix{
({\hat e\over\hat s\hat c}c^\prime-2e_Hg_Es^\prime)^2\cr
({\hat e\over\hat s\hat c}s^\prime+2e_Hg_Ec^\prime)^2\cr
(-{\hat e\over\hat s\hat c}c^\prime+2e_Hg_Es^\prime)
({\hat e\over\hat s\hat c}s^\prime+2e_Hg_Ec^\prime)}\right)\phantom{aaaai}
\nonumber
\end{eqnarray}
\begin{eqnarray}
G^\prime S^0:\phantom{aaaaa}
-4\tilde A(q^2,\hat M_{Z^\prime},\hat M_{S^0})\times\left(\matrix{
e_S^2g_E^2s^{\prime2}\cr
e_S^2g_E^2c^{\prime2}\cr
e_S^2g_E^2c^\prime s^\prime}\right)
\phantom{aaaaaaaaaaaaaaaaaaaaaaaaa}\nonumber
\end{eqnarray}
\begin{eqnarray}
Z^0h^0:\phantom{aaaaa}
+{1\over4}\hat v_H^2b_0(q^2,\hat M_{Z^0},\hat M_{h^0})\times\left(\matrix{
(-{\hat e\over\hat s\hat c}c^\prime+2e_Hg_Es^\prime)^4\cr
(-{\hat e\over\hat s\hat c}c^\prime+2e_Hg_Es^\prime)^2
({\hat e\over\hat s\hat c}s^\prime+2e_Hg_Ec^\prime)^2\cr
(-{\hat e\over\hat s\hat c}c^\prime+2e_Hg_Es^\prime)^3
({\hat e\over\hat s\hat c}s^\prime+2e_Hg_Ec^\prime)}\right)\nonumber
\end{eqnarray}
\begin{eqnarray}
Z^\prime h^0:\phantom{aaaaa}
+{1\over4}\hat v_H^2b_0(q^2,\hat M_{Z^\prime},\hat M_{h^0})
\times\left(\matrix{
(-{\hat e\over\hat s\hat c}c^\prime+2e_Hg_Es^\prime)^2
({\hat e\over\hat s\hat c}s^\prime+2e_Hg_Ec^\prime)^2\cr
({\hat e\over\hat s\hat c}s^\prime+2e_Hg_Ec^\prime)^4\cr
(-{\hat e\over\hat s\hat c}c^\prime+2e_Hg_Es^\prime)
({\hat e\over\hat s\hat c}s^\prime+2e_Hg_Ec^\prime)^3
}\right)\nonumber
\end{eqnarray}
\begin{eqnarray}
Z^0S^0:\phantom{aaaaa}
+4\hat v_S^2e_S^4g_E^4b_0(q^2,\hat M_{Z^0},\hat M_{S^0})\times\left(\matrix{
s^{\prime4}\cr
c^{\prime2}s^{\prime2}\cr
c^\prime s^{\prime3}}\right)\phantom{aaaaaaaaaaaaaaa}\nonumber
\end{eqnarray}
\begin{eqnarray}
Z^\prime S^0:\phantom{aaaaa}
+4\hat v_S^2e_S^4g_E^4b_0(q^2,\hat M_{Z^\prime},\hat M_{S^0})
\times\left(\matrix{
c^{\prime2}s^{\prime2}\cr
c^{\prime4}\cr
c^{\prime3}s^\prime}\right)\phantom{aaaaaaaaaaaaaaa}\nonumber
\end{eqnarray}
To simplify the calculations we have assumed here that the 
calar fields $H$ and $S$ do not mix in the potential, so that the Higgs
boson $h^0$ comes only from the doublet $H$ and $S^0$ only from the 
singlet $S^0$.  

The fermion contribution to $\Pi_{Z_1Z_2}(q^2)$ reads
\begin{eqnarray}
\Pi_{Z^iZ^j}^{\rm ferm}(q^2)
=\sum_fN_c^{(f)}\left\{2(c^{Z^i}_{fL}c^{Z^j}_{fR}
+c^{Z^i}_{fR}c^{Z^j}_{fL})m^2_fb_0(q^2,m_f,m_f)\phantom{aaaaaaaaaaaaaa}
\right.\label{eqn:PIZZfermionic}\\
\left.
+(c^{Z^i}_{fL}c^{Z^j}_{fL}+c^{Z^i}_{fR}c^{Z^j}_{fR})
\left[4\tilde A(q^2,m_f,m_f) 
+(q^2-2m^2_f)b_0(q^2,m_f,m_f)\right]\right\}\nonumber
\end{eqnarray}
where $N_c$ is the colour factor and the couplings $c^{Z^i}_{fL}$, 
$c^{Z^i}_{fR}$ can be read off from (\ref{eqn:Z0couplings}) and
(\ref{eqn:ZPcouplings}).

\section{Loop functions}
\label{app:functions}
\setcounter{equation}{0}

Here we define some loop functions to make the calculations presented
in the text complete.
\begin{eqnarray}
16\pi^2 a(m)= m^2\left(\eta_{\rm div}-1+m^2+\ln{m^2\over\mu^2}\right)
\end{eqnarray}
\begin{eqnarray}
16\pi^2 b_0(q^2,m_1,m_2)=\eta_{\rm div}+\int_0^1dx
\ln{q^2x(x-1)+xm^2_1+(1-x)m^2_2\over\mu^2}
\end{eqnarray}
\begin{eqnarray}
16\pi^2 b_0(0,m_1,m_2)=\eta_{\rm div}-1+{m^2_1\over m^2_1-m^2_2}
\ln{m^2_1\over \mu^2}+{m^2_2\over m^2_2-m^2_1}
\ln{m^2_2\over \mu^2}
\end{eqnarray}
\begin{eqnarray}
\tilde A(q^2,m_1,m_2)&=& -{1\over6}a(m_1)-{1\over6}a(m_2)
+{1\over6}(m_1^2+m_2^2-{q^2\over2})~b_0(q^2,m_1,m_2) \nonumber\\
&+&{m_1^2-m_2^2\over12q^2}\left[a(m_1)-a(m_2)-(m_1^2-m_2^2)~b_0(q^2,m_1,m_2)
\right]\nonumber\\
&-&{1\over16\pi^2}{1\over6}(m_1^2+m_2^2-{q^2\over3})
\end{eqnarray}
The divergent part of $\tilde A(q^2,m_1,m_2)$ is
\begin{eqnarray}
16\pi^2 \left[\tilde A(q^2,m_1,m_2)\right]_{\rm div}
=-{1\over12}q^2\eta_{\rm div}~,
\end{eqnarray}
$\tilde A(0,m_1,m_2)$ is finite and reads
\begin{eqnarray}
16\pi^2 \tilde A(0,m_1,m_2)
=-{1\over8}\left[m_1^2+m_2^2-{2m_1^2m_2^2\over m_1^2-m_2^2}
\log{m_1^2\over m_2^2}\right]\equiv-{1\over8}g(m_1,m_2)
\end{eqnarray}

\end{document}